\documentclass[12pt]{article}
\newcommand{\be}{\begin{equation}}
\newcommand{\ee}{\end{equation}}
\newcommand{\ba}{\begin{eqnarray}}
\newcommand{\ea}{\end{eqnarray}}
\begin{document}
\title{Dimensional reduction, magnetic flux strings, and domain walls}
\author{C.~D.\ Fosco$^a$\thanks{CONICET}\,, A.\
  L\'opez$^a$\thanks{CONICET}\, and\, F.~A.\
  Schaposnik$^b$\thanks{Investigador CICBA, Argentina}
  \\
  {\normalsize\it $^a$Centro At\'omico Bariloche,
    8400 Bariloche, Argentina}\\
  {\normalsize\it
    $^b$Departamento de F\'\i sica, Universidad Nacional de La Plata}\\
  {\normalsize\it C.C. 67, 1900 La Plata, Argentina}}
\maketitle
\begin{abstract}
  We study some consequences of dimensionally reducing systems with
  massless fermions and Abelian gauge fields from $3+1$ to $2+1$
  dimensions. We first consider fermions in the presence of an
  external Abelian gauge field. In the reduced theory, obtained by
  compactifying one of the coordinates `a la Kaluza-Klein, magnetic
  flux strings are mapped into domain wall defects. Fermionic zero
  modes, localized around the flux strings of the $3+1$ dimensional
  theory, become also zero modes in the reduced theory, via the Callan
  and Harvey mechanism, and are concentrated around the domain wall
  defects.  We also study a dynamical model: massless $QED_4$, with
  fermions confined to a plane, deriving the effective action that
  describes the `planar' system.
\end{abstract}
\bigskip
\newpage
\section{Introduction}\label{intro}
Topological defects in quantum field theory~\cite{raja} play an
important role in the description of many interesting phenomena, both
in high energy and condensed matter physics applications. Being part
of the non-perturbative spectrum, the understanding of their
properties demands the study of topological classes of the field
configuration space, and the realization of the associated topological
invariants in the model under consideration.  For some systems
containing fermions, fermionic zero modes arise whenever the
topological charge is different from zero. This effect has many
important consequences for the study of the low momentum effective
theory. Moreover, since they are gapless excitations, they will
strongly affect the response functions of the model~\cite{fl}.

In this paper we are concerned with the relationship between
topological defects and zero modes in theories related by
dimensional reduction. In particular, for the concrete example of
fermions in $3+1$ dimensions coupled to an Abelian gauge field, we
shall derive the low-momentum effective action for the
dimensionally reduced theory.  Being originally a massless theory,
the momentum scale will  be fixed by the inverse of the
compactification length.  Also, as it will be explained, quantum
corrections coming from all the Kaluza-Klein modes have to be
included, to be consistent with large gauge invariance.

We will first consider the case of massless Dirac fermions in the
presence of an external Abelian field. We then go on to consider a
dynamical model: massless $QED$ in $3+1$ dimensions. The reduction is
here introduced for the fermions, assumed to be confined to a planar
spatial region. Being the only sources of the gauge field, this
induces a reduction also for the gauge field, regardless of the fact
that it is {\em apriori\/} unconfined. For this system we derive the
effective theory, showing that there is room for the existence of
`stripe' defects, i.e, localized zero modes that appear due to changes
in the sign of the fermionic mass. This is just a manifestation of the
well known Callan and Harvey mechanism \cite{callan}. When there is
translation invariance along one of the spatial directions, the
dynamics of the defects is governed by a Sine-Gordon like action.
Moreover, the soliton charge is linked to the number of zero modes.

The organization of this paper is as follows: in
section~\ref{free}, we consider fermions in the presence of an
external (non-dynamical) gauge field, in $3+1$ dimensions. To take
into account quantum effects, section~\ref{derexp} deals with the
evaluation of the effective action using a derivative expansion
approximation for the fermionic determinant. In section~\ref{qed},
we derive the low-momentum effective theory for the interacting
case of massless $QED_4$. Finally,  section~\ref{cyd} is devoted
to the conclusions.

\section{`Non-interacting' case}\label{free}
By `non-interacting' we refer to a model containing a fermionic field
$\psi$ in the presence of an external (i.e., non-dynamical) gauge
field $A_\alpha$.  The Euclidean action for the present case is simply
defined by
\begin{equation}\label{defsf}
S_F = \int d^4x \,{\bar\psi} \not \!\! D \psi
\end{equation}
where
\begin{equation}
\not \!\! D \,=\, \gamma_\alpha D_\alpha \;\;\;\;\; D_\alpha =
\partial_\alpha + i e A_\alpha \,.
\end{equation}
We shall adopt the convention that indices from the beginning of the
Greek alphabet always run over the values $0,1,2,3$ while those from
the end belong to the set $0,1,2$ and are reserved for the coordinates
of the reduced spacetime.  The third spatial coordinate, $x_3$, is
compactified: $0 \leq x_3 \leq L$, and shall be sometimes also denoted
`$s$', to emphasize its special role, as opposed to the uncompactified
coordinates, $x_\mu$.

We shall first be concerned with configurations corresponding to
magnetic flux strings in $3+1$ dimensions, with the strings {\em
  entirely contained in the reduced hyperplane\/} $x_3 =
0$~\footnote{This is rather different to the more standard situation
  of a flux tube pointing in the $x_3$ direction, and thus
  corresponding to a magnetic {\em vortex\/} in $2+1$ dimensions.}.
These configurations are here considered regardless of the mechanism
which may create them.
To have these magnetic field structures, special gauge field
configurations are required. Beforehand, we will use part of the gauge
freedom to restrict the external gauge field configurations to the
ones satisfying the condition:
\begin{equation}
\partial_3 A_3 \,=\, 0 \,,
\end{equation}
which can always be fulfilled, by performing a regular gauge
transformation connected to the identity. With this choice, \mbox{$A_3
  (x,s) \,\to\, {\tilde A}_3 (x)$},
\begin{equation}
{\tilde A}_3 (x) \equiv \frac{1}{L} \int_0^L ds A_s (x,s)
\end{equation}

Besides considering static magnetic flux strings, we would also like
to include a `small' gauge field configuration in order to find the
effective action. The strings, being topological objects, should be
treated non-perturbatively, while the small part is assumed to be
intrinsically perturbative.

In $3+1$ dimensions, the magnetic field components $B_i$ ($i=1,2,3$)
are, as usual, defined by $B_i = \epsilon_{ijk} \partial_j A_k$. As
the flux strings lay on the $x_3=0$ plane, the third normal component
of the magnetic field vanishes: $B_3=0$. Moreover, we can generate the
two non-vanishing components $B_1$, $B_2$ using a gauge field such
that $A_1=A_2=0$, and ${\tilde A}_3\neq 0$. With this choice,
\begin{equation}
B_i (x) \,=\, \epsilon_{ij} \partial_j {\tilde A}_3 (x)\;.
\end{equation}

It is evident that, from the point of view of the reduced, planar
theory, ${\tilde A}_3$ is a scalar field under transformations mapping
the plane onto itself. It is convenient to use the definition ${\tilde
  A}_3 (x) \,=\, \varphi (x)$, with $x \equiv (x_0,x_1,x_2)$, to make
explicit the fact that ${\tilde A}_3$ behaves like a scalar.

In the case of a flux string along the $x_1$ axis, the magnetic field
is:
\begin{equation}\label{string}
B_1 \,=\, \xi \,\delta (x_3) \delta (x_2)\;,
\end{equation}
where $\xi$ is a constant~\footnote{This $\delta$-like configuration
  is, of course, an idealization. Realistic configurations will be,
  although highly concentrated, {\em regular\/} functions.}.

We need to write now a gauge field configuration leading to the
magnetic field of (\ref{string}). There are of course many
possibilities but, interestingly enough, if we want to dimensionally
reduce the theory, this freedom is substantially reduced. The reason
is that we do not want a gauge field configuration having a nontrivial
$x_3$ dependence far away from the $x_3=0$ hyperplane. The natural
choice is then a gauge field strongly concentrated on $x_3=0$, and
this {\em determines\/} the gauge field configuration to be of the
form:
\begin{equation}
A_3 (x) \,=\, \frac{\xi}{2} \,\delta (x_3) {\rm sign} (x_2) \;,
\end{equation}
which implies, for the $2+1$ dimensional scalar field
\begin{equation}\label{phisol}
\varphi (x) \,=\,\frac{\xi}{2 L} {\rm sign} (x_2) \;.
\end{equation}

The constant $\xi$ is, of course, related to the (quantized) total
flux of the string, thus $\xi = \frac{2 \pi k}{e}$, where $k$ is an
integer, counting the number of elementary fluxes.

We note the emergence of a step-like configuration for the scalar
field. This scalar field, as we shall see next, plays the role of the
mass for $2+1$ dimensional fermions. Hence, invoking the Callan and
Harvey mechanism~\cite{callan}, we see that there are zero modes
concentrated on the walls, which indeed correspond to the
dimensionally reduced version of the zero modes around the flux
strings. It is worth noting that the scalar field profile is
discontinuous because of the assumed $\delta$-like profile for the
magnetic field. It is convenient to use instead a regular profile,
\begin{equation}\label{phisol2}
\varphi(x) \,=\, \frac{\xi}{2L} \, h_\Delta (x_2)
\end{equation}
where $h_\Delta (x_2)$ is a smoothed version of the step function,
for example
\begin{equation}
h_\Delta (x_2)\,=\,{\rm tanh} (\frac{x_2}{\Delta})
\end{equation}
where $\Delta$ is a length which essentially measures the `width' of
the stripe.

To discuss issues related to the fermionic zero modes, like the
relationship between the number of modes in $3+1$ and $2+1$
dimensions, a more detailed study of the fermionic action
(\ref{defsf}) is required. The full gauge field configuration
(perturbative plus non-perturbative) is assumed to be $s$-independent.
This simplifying assumption allows us to take advantage of the fact
that the gauge field ${\tilde A}_3$ is independent of $s=x_3$ to
Fourier expand the fermionic fields:
\begin{eqnarray}
\psi(x,x_3) &=& \frac{1}{\sqrt L} \sum_{n=-\infty}^{+\infty}
e^{i \omega_n x_3} \psi_n (x) \nonumber\\
{\bar\psi}(x,x_3) &=& \frac{1}{\sqrt L} \sum_{n=-\infty}^{+\infty}
e^{-i \omega_n x_3} {\bar\psi}_n (x) \;,
\end{eqnarray}
with $\omega_n={2\pi n}/{L}\,$. We then obtain for $S_F$ a
representation as a series of decoupled $2+1$ dimensional actions
\begin{equation}\label{decomp}
S_F \,=\, \sum_{n=-\infty}^{+\infty} S_{F,n} \;.
\end{equation}
The explicit form for the action $S_{F,n}$ corresponding to each
Fourier mode is:
\begin{equation}
S_{F,n}\,=\, \int d^3x \,{\bar\psi}_n (x)
[\gamma_\mu D_\mu + i \gamma_3 (\omega_n + e \varphi)]
\,\psi_n (x) \;.
\end{equation}
We assume the Dirac matrices are in the representation:
\begin{equation}
\gamma_0 \,=\,
\left(\begin{array}{cc}
0 & \sigma_3\\
\sigma_3 & 0
\end{array}\right)
\,
\gamma_1 \,=\,
\left(\begin{array}{cc}
0 & \sigma_1\\
\sigma_1 & 0
\end{array}\right)
\,
\gamma_2 \,=\,
\left(\begin{array}{cc}
0 & \sigma_2\\
\sigma_2 & 0
\end{array}\right)
\,
\gamma_3 \,=\,
\left(\begin{array}{cc}
0 & i I\\
-i I & 0
\end{array}\right)
\end{equation}
where $\sigma_j$, with $j=1,2,3$, denote the usual Pauli matrices, and
$I$ is the $2 \times 2$ identity matrix. Writing each four-component
fermionic field $\psi_n$ in terms of two $2$-component fermions
$\chi_n^{(a)}$, ($a=1,2$), we see that
\begin{equation}
S_{F,n}=\int d^3x \, \left[
{\bar\chi}^{(1)}_n (\not \! d + \omega_n+e \varphi ) \chi^{(1)}_n \,
+\,{\bar\chi}^{(2)}_n(x)
(\not \! d - \omega_n-e \varphi ) \chi^{(2)}_n(x) \right] \;.
\end{equation}
We have introduced the notation $\not \! d$ to refer to the Dirac
operator in $2+1$ dimensions, acting on $2$-component fermions. More
explicitly,
\begin{equation}\label{defd}
\not \! d \,=\, \sigma_\mu D_\mu
\end{equation}
with $\sigma_0 \equiv \sigma_3$, and $\sigma_j$ ($j = 1,2$) again
denoting the usual Pauli matrices. The covariant derivative depends on
a gauge field which, in our approximation, is a function of $x_\mu$
only.

It is important to realize that all the fermionic modes should be kept
if {\em large\/} gauge invariance is to be preserved, since large
gauge invariance amounts to shifts in $\varphi$, and this field is
directly coupled to the fermionic field.

It is now easy to study the Callan and Harvey fermionic zero modes of
the reduced theory, assuming that in the original model in $3+1$
dimensions the background produced a non-vanishing, quantized,
magnetic flux. We first realize that the decomposition (\ref{decomp})
shows that there is an infinite number of $2+1$ dimensional Dirac
fermions, each Fourier mode consisting of two flavours, distinguished
by the sign of their mass terms. A Callan and Harvey zero mode for the
Fourier component $n$ shall appear whenever $\omega_n + e \varphi (x)$
crosses zero more or less sharply. Thus, the configuration
(\ref{phisol}) (in its smooth version (\ref{phisol2})) shows that
there is at least a zero mode for the $n=0$ component, but, if there
is a flux string with charge $k$ , the scalar field profile will cross
as many values of $\omega_n$, implying that there are, indeed, $k$
zero modes in the reduced theory. To see this, we take into account
the expression (\ref{phisol2}) for the domain wall configuration. The
equation determining the locii of the zero modes may be written as
\begin{equation}
h_\Delta (x_2) \,=\, \frac{2}{k} \times {\rm integer} \;,
\end{equation}
where $k$ is the (fixed) number of elementary fluxes. In Figure 1 we
show, as an example, the case of a stripe configuration $h_\Delta$
with $k=9$. The horizontal lines correspond to different $n$'s, and a
zero mode is produced whenever one of these lines intersects the
curve.

\setlength{\unitlength}{0.240900pt}
\ifx\plotpoint\undefined\newsavebox{\plotpoint}\fi
\begin{picture}(1500,900)(0,0)
\font\gnuplot=cmr10 at 10pt
\gnuplot
\sbox{\plotpoint}{\rule[-0.200pt]{0.400pt}{0.400pt}}%
\put(140.0,82.0){\rule[-0.200pt]{4.818pt}{0.400pt}}
\put(120,82){\makebox(0,0)[r]{-1}}
\put(1419.0,82.0){\rule[-0.200pt]{4.818pt}{0.400pt}}
\put(140.0,160.0){\rule[-0.200pt]{4.818pt}{0.400pt}}
\put(120,160){\makebox(0,0)[r]{-0.8}}
\put(1419.0,160.0){\rule[-0.200pt]{4.818pt}{0.400pt}}
\put(140.0,238.0){\rule[-0.200pt]{4.818pt}{0.400pt}}
\put(120,238){\makebox(0,0)[r]{-0.6}}
\put(1419.0,238.0){\rule[-0.200pt]{4.818pt}{0.400pt}}
\put(140.0,315.0){\rule[-0.200pt]{4.818pt}{0.400pt}}
\put(120,315){\makebox(0,0)[r]{-0.4}}
\put(1419.0,315.0){\rule[-0.200pt]{4.818pt}{0.400pt}}
\put(140.0,393.0){\rule[-0.200pt]{4.818pt}{0.400pt}}
\put(120,393){\makebox(0,0)[r]{-0.2}}
\put(1419.0,393.0){\rule[-0.200pt]{4.818pt}{0.400pt}}
\put(140.0,471.0){\rule[-0.200pt]{4.818pt}{0.400pt}}
\put(120,471){\makebox(0,0)[r]{0}}
\put(1419.0,471.0){\rule[-0.200pt]{4.818pt}{0.400pt}}
\put(140.0,549.0){\rule[-0.200pt]{4.818pt}{0.400pt}}
\put(120,549){\makebox(0,0)[r]{0.2}}
\put(1419.0,549.0){\rule[-0.200pt]{4.818pt}{0.400pt}}
\put(140.0,627.0){\rule[-0.200pt]{4.818pt}{0.400pt}}
\put(120,627){\makebox(0,0)[r]{0.4}}
\put(1419.0,627.0){\rule[-0.200pt]{4.818pt}{0.400pt}}
\put(140.0,704.0){\rule[-0.200pt]{4.818pt}{0.400pt}}
\put(120,704){\makebox(0,0)[r]{0.6}}
\put(1419.0,704.0){\rule[-0.200pt]{4.818pt}{0.400pt}}
\put(140.0,782.0){\rule[-0.200pt]{4.818pt}{0.400pt}}
\put(120,782){\makebox(0,0)[r]{0.8}}
\put(1419.0,782.0){\rule[-0.200pt]{4.818pt}{0.400pt}}
\put(140.0,860.0){\rule[-0.200pt]{4.818pt}{0.400pt}}
\put(120,860){\makebox(0,0)[r]{1}}
\put(1419.0,860.0){\rule[-0.200pt]{4.818pt}{0.400pt}}
\put(140.0,82.0){\rule[-0.200pt]{0.400pt}{4.818pt}}
\put(140,41){\makebox(0,0){-10}}
\put(140.0,840.0){\rule[-0.200pt]{0.400pt}{4.818pt}}
\put(465.0,82.0){\rule[-0.200pt]{0.400pt}{4.818pt}}
\put(465,41){\makebox(0,0){-5}}
\put(465.0,840.0){\rule[-0.200pt]{0.400pt}{4.818pt}}
\put(790.0,82.0){\rule[-0.200pt]{0.400pt}{4.818pt}}
\put(790,41){\makebox(0,0){0}}
\put(790.0,840.0){\rule[-0.200pt]{0.400pt}{4.818pt}}
\put(1114.0,82.0){\rule[-0.200pt]{0.400pt}{4.818pt}}
\put(1114,41){\makebox(0,0){5}}
\put(1114.0,840.0){\rule[-0.200pt]{0.400pt}{4.818pt}}
\put(1439.0,82.0){\rule[-0.200pt]{0.400pt}{4.818pt}}
\put(1439,41){\makebox(0,0){10}}
\put(1439.0,840.0){\rule[-0.200pt]{0.400pt}{4.818pt}}
\put(140.0,82.0){\rule[-0.200pt]{312.929pt}{0.400pt}}
\put(1439.0,82.0){\rule[-0.200pt]{0.400pt}{187.420pt}}
\put(140.0,860.0){\rule[-0.200pt]{312.929pt}{0.400pt}}
\put(140.0,82.0){\rule[-0.200pt]{0.400pt}{187.420pt}}
\put(140,82){\usebox{\plotpoint}}
\put(547,81.67){\rule{3.132pt}{0.400pt}}
\multiput(547.00,81.17)(6.500,1.000){2}{\rule{1.566pt}{0.400pt}}
\put(140.0,82.0){\rule[-0.200pt]{98.046pt}{0.400pt}}
\put(586,82.67){\rule{3.132pt}{0.400pt}}
\multiput(586.00,82.17)(6.500,1.000){2}{\rule{1.566pt}{0.400pt}}
\put(599,83.67){\rule{3.132pt}{0.400pt}}
\multiput(599.00,83.17)(6.500,1.000){2}{\rule{1.566pt}{0.400pt}}
\put(612,85.17){\rule{2.700pt}{0.400pt}}
\multiput(612.00,84.17)(7.396,2.000){2}{\rule{1.350pt}{0.400pt}}
\put(625,87.17){\rule{2.900pt}{0.400pt}}
\multiput(625.00,86.17)(7.981,2.000){2}{\rule{1.450pt}{0.400pt}}
\multiput(639.00,89.60)(1.797,0.468){5}{\rule{1.400pt}{0.113pt}}
\multiput(639.00,88.17)(10.094,4.000){2}{\rule{0.700pt}{0.400pt}}
\multiput(652.00,93.59)(1.378,0.477){7}{\rule{1.140pt}{0.115pt}}
\multiput(652.00,92.17)(10.634,5.000){2}{\rule{0.570pt}{0.400pt}}
\multiput(665.00,98.59)(0.824,0.488){13}{\rule{0.750pt}{0.117pt}}
\multiput(665.00,97.17)(11.443,8.000){2}{\rule{0.375pt}{0.400pt}}
\multiput(678.00,106.58)(0.539,0.492){21}{\rule{0.533pt}{0.119pt}}
\multiput(678.00,105.17)(11.893,12.000){2}{\rule{0.267pt}{0.400pt}}
\multiput(691.58,118.00)(0.493,0.616){23}{\rule{0.119pt}{0.592pt}}
\multiput(690.17,118.00)(13.000,14.771){2}{\rule{0.400pt}{0.296pt}}
\multiput(704.58,134.00)(0.493,0.933){23}{\rule{0.119pt}{0.838pt}}
\multiput(703.17,134.00)(13.000,22.260){2}{\rule{0.400pt}{0.419pt}}
\multiput(717.58,158.00)(0.493,1.290){23}{\rule{0.119pt}{1.115pt}}
\multiput(716.17,158.00)(13.000,30.685){2}{\rule{0.400pt}{0.558pt}}
\multiput(730.58,191.00)(0.494,1.562){25}{\rule{0.119pt}{1.329pt}}
\multiput(729.17,191.00)(14.000,40.242){2}{\rule{0.400pt}{0.664pt}}
\multiput(744.58,234.00)(0.493,2.201){23}{\rule{0.119pt}{1.823pt}}
\multiput(743.17,234.00)(13.000,52.216){2}{\rule{0.400pt}{0.912pt}}
\multiput(757.58,290.00)(0.493,2.638){23}{\rule{0.119pt}{2.162pt}}
\multiput(756.17,290.00)(13.000,62.514){2}{\rule{0.400pt}{1.081pt}}
\multiput(770.58,357.00)(0.493,2.955){23}{\rule{0.119pt}{2.408pt}}
\multiput(769.17,357.00)(13.000,70.003){2}{\rule{0.400pt}{1.204pt}}
\multiput(783.58,432.00)(0.493,3.074){23}{\rule{0.119pt}{2.500pt}}
\multiput(782.17,432.00)(13.000,72.811){2}{\rule{0.400pt}{1.250pt}}
\multiput(796.58,510.00)(0.493,2.955){23}{\rule{0.119pt}{2.408pt}}
\multiput(795.17,510.00)(13.000,70.003){2}{\rule{0.400pt}{1.204pt}}
\multiput(809.58,585.00)(0.493,2.638){23}{\rule{0.119pt}{2.162pt}}
\multiput(808.17,585.00)(13.000,62.514){2}{\rule{0.400pt}{1.081pt}}
\multiput(822.58,652.00)(0.493,2.201){23}{\rule{0.119pt}{1.823pt}}
\multiput(821.17,652.00)(13.000,52.216){2}{\rule{0.400pt}{0.912pt}}
\multiput(835.58,708.00)(0.494,1.562){25}{\rule{0.119pt}{1.329pt}}
\multiput(834.17,708.00)(14.000,40.242){2}{\rule{0.400pt}{0.664pt}}
\multiput(849.58,751.00)(0.493,1.290){23}{\rule{0.119pt}{1.115pt}}
\multiput(848.17,751.00)(13.000,30.685){2}{\rule{0.400pt}{0.558pt}}
\multiput(862.58,784.00)(0.493,0.933){23}{\rule{0.119pt}{0.838pt}}
\multiput(861.17,784.00)(13.000,22.260){2}{\rule{0.400pt}{0.419pt}}
\multiput(875.58,808.00)(0.493,0.616){23}{\rule{0.119pt}{0.592pt}}
\multiput(874.17,808.00)(13.000,14.771){2}{\rule{0.400pt}{0.296pt}}
\multiput(888.00,824.58)(0.539,0.492){21}{\rule{0.533pt}{0.119pt}}
\multiput(888.00,823.17)(11.893,12.000){2}{\rule{0.267pt}{0.400pt}}
\multiput(901.00,836.59)(0.824,0.488){13}{\rule{0.750pt}{0.117pt}}
\multiput(901.00,835.17)(11.443,8.000){2}{\rule{0.375pt}{0.400pt}}
\multiput(914.00,844.59)(1.378,0.477){7}{\rule{1.140pt}{0.115pt}}
\multiput(914.00,843.17)(10.634,5.000){2}{\rule{0.570pt}{0.400pt}}
\multiput(927.00,849.60)(1.797,0.468){5}{\rule{1.400pt}{0.113pt}}
\multiput(927.00,848.17)(10.094,4.000){2}{\rule{0.700pt}{0.400pt}}
\put(940,853.17){\rule{2.900pt}{0.400pt}}
\multiput(940.00,852.17)(7.981,2.000){2}{\rule{1.450pt}{0.400pt}}
\put(954,855.17){\rule{2.700pt}{0.400pt}}
\multiput(954.00,854.17)(7.396,2.000){2}{\rule{1.350pt}{0.400pt}}
\put(967,856.67){\rule{3.132pt}{0.400pt}}
\multiput(967.00,856.17)(6.500,1.000){2}{\rule{1.566pt}{0.400pt}}
\put(980,857.67){\rule{3.132pt}{0.400pt}}
\multiput(980.00,857.17)(6.500,1.000){2}{\rule{1.566pt}{0.400pt}}
\put(560.0,83.0){\rule[-0.200pt]{6.263pt}{0.400pt}}
\put(1019,858.67){\rule{3.132pt}{0.400pt}}
\multiput(1019.00,858.17)(6.500,1.000){2}{\rule{1.566pt}{0.400pt}}
\put(993.0,859.0){\rule[-0.200pt]{6.263pt}{0.400pt}}
\put(1032.0,860.0){\rule[-0.200pt]{98.046pt}{0.400pt}}
\put(140,125){\usebox{\plotpoint}}
\put(140.00,125.00){\usebox{\plotpoint}}
\put(160.76,125.00){\usebox{\plotpoint}}
\put(181.51,125.00){\usebox{\plotpoint}}
\put(202.27,125.00){\usebox{\plotpoint}}
\put(223.02,125.00){\usebox{\plotpoint}}
\put(243.78,125.00){\usebox{\plotpoint}}
\put(264.53,125.00){\usebox{\plotpoint}}
\put(285.29,125.00){\usebox{\plotpoint}}
\put(306.04,125.00){\usebox{\plotpoint}}
\put(326.80,125.00){\usebox{\plotpoint}}
\put(347.55,125.00){\usebox{\plotpoint}}
\put(368.31,125.00){\usebox{\plotpoint}}
\put(389.07,125.00){\usebox{\plotpoint}}
\put(409.82,125.00){\usebox{\plotpoint}}
\put(430.58,125.00){\usebox{\plotpoint}}
\put(451.33,125.00){\usebox{\plotpoint}}
\put(472.09,125.00){\usebox{\plotpoint}}
\put(492.84,125.00){\usebox{\plotpoint}}
\put(513.60,125.00){\usebox{\plotpoint}}
\put(534.35,125.00){\usebox{\plotpoint}}
\put(555.11,125.00){\usebox{\plotpoint}}
\put(575.87,125.00){\usebox{\plotpoint}}
\put(596.62,125.00){\usebox{\plotpoint}}
\put(617.38,125.00){\usebox{\plotpoint}}
\put(638.13,125.00){\usebox{\plotpoint}}
\put(658.89,125.00){\usebox{\plotpoint}}
\put(679.64,125.00){\usebox{\plotpoint}}
\put(700.40,125.00){\usebox{\plotpoint}}
\put(721.15,125.00){\usebox{\plotpoint}}
\put(741.91,125.00){\usebox{\plotpoint}}
\put(762.66,125.00){\usebox{\plotpoint}}
\put(783.42,125.00){\usebox{\plotpoint}}
\put(804.18,125.00){\usebox{\plotpoint}}
\put(824.93,125.00){\usebox{\plotpoint}}
\put(845.69,125.00){\usebox{\plotpoint}}
\put(866.44,125.00){\usebox{\plotpoint}}
\put(887.20,125.00){\usebox{\plotpoint}}
\put(907.95,125.00){\usebox{\plotpoint}}
\put(928.71,125.00){\usebox{\plotpoint}}
\put(949.46,125.00){\usebox{\plotpoint}}
\put(970.22,125.00){\usebox{\plotpoint}}
\put(990.98,125.00){\usebox{\plotpoint}}
\put(1011.73,125.00){\usebox{\plotpoint}}
\put(1032.49,125.00){\usebox{\plotpoint}}
\put(1053.24,125.00){\usebox{\plotpoint}}
\put(1074.00,125.00){\usebox{\plotpoint}}
\put(1094.75,125.00){\usebox{\plotpoint}}
\put(1115.51,125.00){\usebox{\plotpoint}}
\put(1136.26,125.00){\usebox{\plotpoint}}
\put(1157.02,125.00){\usebox{\plotpoint}}
\put(1177.77,125.00){\usebox{\plotpoint}}
\put(1198.53,125.00){\usebox{\plotpoint}}
\put(1219.29,125.00){\usebox{\plotpoint}}
\put(1240.04,125.00){\usebox{\plotpoint}}
\put(1260.80,125.00){\usebox{\plotpoint}}
\put(1281.55,125.00){\usebox{\plotpoint}}
\put(1302.31,125.00){\usebox{\plotpoint}}
\put(1323.06,125.00){\usebox{\plotpoint}}
\put(1343.82,125.00){\usebox{\plotpoint}}
\put(1364.57,125.00){\usebox{\plotpoint}}
\put(1385.33,125.00){\usebox{\plotpoint}}
\put(1406.09,125.00){\usebox{\plotpoint}}
\put(1426.84,125.00){\usebox{\plotpoint}}
\put(1439,125){\usebox{\plotpoint}}
\put(140,212){\usebox{\plotpoint}}
\put(140.00,212.00){\usebox{\plotpoint}}
\put(160.76,212.00){\usebox{\plotpoint}}
\put(181.51,212.00){\usebox{\plotpoint}}
\put(202.27,212.00){\usebox{\plotpoint}}
\put(223.02,212.00){\usebox{\plotpoint}}
\put(243.78,212.00){\usebox{\plotpoint}}
\put(264.53,212.00){\usebox{\plotpoint}}
\put(285.29,212.00){\usebox{\plotpoint}}
\put(306.04,212.00){\usebox{\plotpoint}}
\put(326.80,212.00){\usebox{\plotpoint}}
\put(347.55,212.00){\usebox{\plotpoint}}
\put(368.31,212.00){\usebox{\plotpoint}}
\put(389.07,212.00){\usebox{\plotpoint}}
\put(409.82,212.00){\usebox{\plotpoint}}
\put(430.58,212.00){\usebox{\plotpoint}}
\put(451.33,212.00){\usebox{\plotpoint}}
\put(472.09,212.00){\usebox{\plotpoint}}
\put(492.84,212.00){\usebox{\plotpoint}}
\put(513.60,212.00){\usebox{\plotpoint}}
\put(534.35,212.00){\usebox{\plotpoint}}
\put(555.11,212.00){\usebox{\plotpoint}}
\put(575.87,212.00){\usebox{\plotpoint}}
\put(596.62,212.00){\usebox{\plotpoint}}
\put(617.38,212.00){\usebox{\plotpoint}}
\put(638.13,212.00){\usebox{\plotpoint}}
\put(658.89,212.00){\usebox{\plotpoint}}
\put(679.64,212.00){\usebox{\plotpoint}}
\put(700.40,212.00){\usebox{\plotpoint}}
\put(721.15,212.00){\usebox{\plotpoint}}
\put(741.91,212.00){\usebox{\plotpoint}}
\put(762.66,212.00){\usebox{\plotpoint}}
\put(783.42,212.00){\usebox{\plotpoint}}
\put(804.18,212.00){\usebox{\plotpoint}}
\put(824.93,212.00){\usebox{\plotpoint}}
\put(845.69,212.00){\usebox{\plotpoint}}
\put(866.44,212.00){\usebox{\plotpoint}}
\put(887.20,212.00){\usebox{\plotpoint}}
\put(907.95,212.00){\usebox{\plotpoint}}
\put(928.71,212.00){\usebox{\plotpoint}}
\put(949.46,212.00){\usebox{\plotpoint}}
\put(970.22,212.00){\usebox{\plotpoint}}
\put(990.98,212.00){\usebox{\plotpoint}}
\put(1011.73,212.00){\usebox{\plotpoint}}
\put(1032.49,212.00){\usebox{\plotpoint}}
\put(1053.24,212.00){\usebox{\plotpoint}}
\put(1074.00,212.00){\usebox{\plotpoint}}
\put(1094.75,212.00){\usebox{\plotpoint}}
\put(1115.51,212.00){\usebox{\plotpoint}}
\put(1136.26,212.00){\usebox{\plotpoint}}
\put(1157.02,212.00){\usebox{\plotpoint}}
\put(1177.77,212.00){\usebox{\plotpoint}}
\put(1198.53,212.00){\usebox{\plotpoint}}
\put(1219.29,212.00){\usebox{\plotpoint}}
\put(1240.04,212.00){\usebox{\plotpoint}}
\put(1260.80,212.00){\usebox{\plotpoint}}
\put(1281.55,212.00){\usebox{\plotpoint}}
\put(1302.31,212.00){\usebox{\plotpoint}}
\put(1323.06,212.00){\usebox{\plotpoint}}
\put(1343.82,212.00){\usebox{\plotpoint}}
\put(1364.57,212.00){\usebox{\plotpoint}}
\put(1385.33,212.00){\usebox{\plotpoint}}
\put(1406.09,212.00){\usebox{\plotpoint}}
\put(1426.84,212.00){\usebox{\plotpoint}}
\put(1439,212){\usebox{\plotpoint}}
\put(140,298){\usebox{\plotpoint}}
\put(140.00,298.00){\usebox{\plotpoint}}
\put(160.76,298.00){\usebox{\plotpoint}}
\put(181.51,298.00){\usebox{\plotpoint}}
\put(202.27,298.00){\usebox{\plotpoint}}
\put(223.02,298.00){\usebox{\plotpoint}}
\put(243.78,298.00){\usebox{\plotpoint}}
\put(264.53,298.00){\usebox{\plotpoint}}
\put(285.29,298.00){\usebox{\plotpoint}}
\put(306.04,298.00){\usebox{\plotpoint}}
\put(326.80,298.00){\usebox{\plotpoint}}
\put(347.55,298.00){\usebox{\plotpoint}}
\put(368.31,298.00){\usebox{\plotpoint}}
\put(389.07,298.00){\usebox{\plotpoint}}
\put(409.82,298.00){\usebox{\plotpoint}}
\put(430.58,298.00){\usebox{\plotpoint}}
\put(451.33,298.00){\usebox{\plotpoint}}
\put(472.09,298.00){\usebox{\plotpoint}}
\put(492.84,298.00){\usebox{\plotpoint}}
\put(513.60,298.00){\usebox{\plotpoint}}
\put(534.35,298.00){\usebox{\plotpoint}}
\put(555.11,298.00){\usebox{\plotpoint}}
\put(575.87,298.00){\usebox{\plotpoint}}
\put(596.62,298.00){\usebox{\plotpoint}}
\put(617.38,298.00){\usebox{\plotpoint}}
\put(638.13,298.00){\usebox{\plotpoint}}
\put(658.89,298.00){\usebox{\plotpoint}}
\put(679.64,298.00){\usebox{\plotpoint}}
\put(700.40,298.00){\usebox{\plotpoint}}
\put(721.15,298.00){\usebox{\plotpoint}}
\put(741.91,298.00){\usebox{\plotpoint}}
\put(762.66,298.00){\usebox{\plotpoint}}
\put(783.42,298.00){\usebox{\plotpoint}}
\put(804.18,298.00){\usebox{\plotpoint}}
\put(824.93,298.00){\usebox{\plotpoint}}
\put(845.69,298.00){\usebox{\plotpoint}}
\put(866.44,298.00){\usebox{\plotpoint}}
\put(887.20,298.00){\usebox{\plotpoint}}
\put(907.95,298.00){\usebox{\plotpoint}}
\put(928.71,298.00){\usebox{\plotpoint}}
\put(949.46,298.00){\usebox{\plotpoint}}
\put(970.22,298.00){\usebox{\plotpoint}}
\put(990.98,298.00){\usebox{\plotpoint}}
\put(1011.73,298.00){\usebox{\plotpoint}}
\put(1032.49,298.00){\usebox{\plotpoint}}
\put(1053.24,298.00){\usebox{\plotpoint}}
\put(1074.00,298.00){\usebox{\plotpoint}}
\put(1094.75,298.00){\usebox{\plotpoint}}
\put(1115.51,298.00){\usebox{\plotpoint}}
\put(1136.26,298.00){\usebox{\plotpoint}}
\put(1157.02,298.00){\usebox{\plotpoint}}
\put(1177.77,298.00){\usebox{\plotpoint}}
\put(1198.53,298.00){\usebox{\plotpoint}}
\put(1219.29,298.00){\usebox{\plotpoint}}
\put(1240.04,298.00){\usebox{\plotpoint}}
\put(1260.80,298.00){\usebox{\plotpoint}}
\put(1281.55,298.00){\usebox{\plotpoint}}
\put(1302.31,298.00){\usebox{\plotpoint}}
\put(1323.06,298.00){\usebox{\plotpoint}}
\put(1343.82,298.00){\usebox{\plotpoint}}
\put(1364.57,298.00){\usebox{\plotpoint}}
\put(1385.33,298.00){\usebox{\plotpoint}}
\put(1406.09,298.00){\usebox{\plotpoint}}
\put(1426.84,298.00){\usebox{\plotpoint}}
\put(1439,298){\usebox{\plotpoint}}
\put(140,385){\usebox{\plotpoint}}
\put(140.00,385.00){\usebox{\plotpoint}}
\put(160.76,385.00){\usebox{\plotpoint}}
\put(181.51,385.00){\usebox{\plotpoint}}
\put(202.27,385.00){\usebox{\plotpoint}}
\put(223.02,385.00){\usebox{\plotpoint}}
\put(243.78,385.00){\usebox{\plotpoint}}
\put(264.53,385.00){\usebox{\plotpoint}}
\put(285.29,385.00){\usebox{\plotpoint}}
\put(306.04,385.00){\usebox{\plotpoint}}
\put(326.80,385.00){\usebox{\plotpoint}}
\put(347.55,385.00){\usebox{\plotpoint}}
\put(368.31,385.00){\usebox{\plotpoint}}
\put(389.07,385.00){\usebox{\plotpoint}}
\put(409.82,385.00){\usebox{\plotpoint}}
\put(430.58,385.00){\usebox{\plotpoint}}
\put(451.33,385.00){\usebox{\plotpoint}}
\put(472.09,385.00){\usebox{\plotpoint}}
\put(492.84,385.00){\usebox{\plotpoint}}
\put(513.60,385.00){\usebox{\plotpoint}}
\put(534.35,385.00){\usebox{\plotpoint}}
\put(555.11,385.00){\usebox{\plotpoint}}
\put(575.87,385.00){\usebox{\plotpoint}}
\put(596.62,385.00){\usebox{\plotpoint}}
\put(617.38,385.00){\usebox{\plotpoint}}
\put(638.13,385.00){\usebox{\plotpoint}}
\put(658.89,385.00){\usebox{\plotpoint}}
\put(679.64,385.00){\usebox{\plotpoint}}
\put(700.40,385.00){\usebox{\plotpoint}}
\put(721.15,385.00){\usebox{\plotpoint}}
\put(741.91,385.00){\usebox{\plotpoint}}
\put(762.66,385.00){\usebox{\plotpoint}}
\put(783.42,385.00){\usebox{\plotpoint}}
\put(804.18,385.00){\usebox{\plotpoint}}
\put(824.93,385.00){\usebox{\plotpoint}}
\put(845.69,385.00){\usebox{\plotpoint}}
\put(866.44,385.00){\usebox{\plotpoint}}
\put(887.20,385.00){\usebox{\plotpoint}}
\put(907.95,385.00){\usebox{\plotpoint}}
\put(928.71,385.00){\usebox{\plotpoint}}
\put(949.46,385.00){\usebox{\plotpoint}}
\put(970.22,385.00){\usebox{\plotpoint}}
\put(990.98,385.00){\usebox{\plotpoint}}
\put(1011.73,385.00){\usebox{\plotpoint}}
\put(1032.49,385.00){\usebox{\plotpoint}}
\put(1053.24,385.00){\usebox{\plotpoint}}
\put(1074.00,385.00){\usebox{\plotpoint}}
\put(1094.75,385.00){\usebox{\plotpoint}}
\put(1115.51,385.00){\usebox{\plotpoint}}
\put(1136.26,385.00){\usebox{\plotpoint}}
\put(1157.02,385.00){\usebox{\plotpoint}}
\put(1177.77,385.00){\usebox{\plotpoint}}
\put(1198.53,385.00){\usebox{\plotpoint}}
\put(1219.29,385.00){\usebox{\plotpoint}}
\put(1240.04,385.00){\usebox{\plotpoint}}
\put(1260.80,385.00){\usebox{\plotpoint}}
\put(1281.55,385.00){\usebox{\plotpoint}}
\put(1302.31,385.00){\usebox{\plotpoint}}
\put(1323.06,385.00){\usebox{\plotpoint}}
\put(1343.82,385.00){\usebox{\plotpoint}}
\put(1364.57,385.00){\usebox{\plotpoint}}
\put(1385.33,385.00){\usebox{\plotpoint}}
\put(1406.09,385.00){\usebox{\plotpoint}}
\put(1426.84,385.00){\usebox{\plotpoint}}
\put(1439,385){\usebox{\plotpoint}}
\put(140,471){\usebox{\plotpoint}}
\put(140.00,471.00){\usebox{\plotpoint}}
\put(160.76,471.00){\usebox{\plotpoint}}
\put(181.51,471.00){\usebox{\plotpoint}}
\put(202.27,471.00){\usebox{\plotpoint}}
\put(223.02,471.00){\usebox{\plotpoint}}
\put(243.78,471.00){\usebox{\plotpoint}}
\put(264.53,471.00){\usebox{\plotpoint}}
\put(285.29,471.00){\usebox{\plotpoint}}
\put(306.04,471.00){\usebox{\plotpoint}}
\put(326.80,471.00){\usebox{\plotpoint}}
\put(347.55,471.00){\usebox{\plotpoint}}
\put(368.31,471.00){\usebox{\plotpoint}}
\put(389.07,471.00){\usebox{\plotpoint}}
\put(409.82,471.00){\usebox{\plotpoint}}
\put(430.58,471.00){\usebox{\plotpoint}}
\put(451.33,471.00){\usebox{\plotpoint}}
\put(472.09,471.00){\usebox{\plotpoint}}
\put(492.84,471.00){\usebox{\plotpoint}}
\put(513.60,471.00){\usebox{\plotpoint}}
\put(534.35,471.00){\usebox{\plotpoint}}
\put(555.11,471.00){\usebox{\plotpoint}}
\put(575.87,471.00){\usebox{\plotpoint}}
\put(596.62,471.00){\usebox{\plotpoint}}
\put(617.38,471.00){\usebox{\plotpoint}}
\put(638.13,471.00){\usebox{\plotpoint}}
\put(658.89,471.00){\usebox{\plotpoint}}
\put(679.64,471.00){\usebox{\plotpoint}}
\put(700.40,471.00){\usebox{\plotpoint}}
\put(721.15,471.00){\usebox{\plotpoint}}
\put(741.91,471.00){\usebox{\plotpoint}}
\put(762.66,471.00){\usebox{\plotpoint}}
\put(783.42,471.00){\usebox{\plotpoint}}
\put(804.18,471.00){\usebox{\plotpoint}}
\put(824.93,471.00){\usebox{\plotpoint}}
\put(845.69,471.00){\usebox{\plotpoint}}
\put(866.44,471.00){\usebox{\plotpoint}}
\put(887.20,471.00){\usebox{\plotpoint}}
\put(907.95,471.00){\usebox{\plotpoint}}
\put(928.71,471.00){\usebox{\plotpoint}}
\put(949.46,471.00){\usebox{\plotpoint}}
\put(970.22,471.00){\usebox{\plotpoint}}
\put(990.98,471.00){\usebox{\plotpoint}}
\put(1011.73,471.00){\usebox{\plotpoint}}
\put(1032.49,471.00){\usebox{\plotpoint}}
\put(1053.24,471.00){\usebox{\plotpoint}}
\put(1074.00,471.00){\usebox{\plotpoint}}
\put(1094.75,471.00){\usebox{\plotpoint}}
\put(1115.51,471.00){\usebox{\plotpoint}}
\put(1136.26,471.00){\usebox{\plotpoint}}
\put(1157.02,471.00){\usebox{\plotpoint}}
\put(1177.77,471.00){\usebox{\plotpoint}}
\put(1198.53,471.00){\usebox{\plotpoint}}
\put(1219.29,471.00){\usebox{\plotpoint}}
\put(1240.04,471.00){\usebox{\plotpoint}}
\put(1260.80,471.00){\usebox{\plotpoint}}
\put(1281.55,471.00){\usebox{\plotpoint}}
\put(1302.31,471.00){\usebox{\plotpoint}}
\put(1323.06,471.00){\usebox{\plotpoint}}
\put(1343.82,471.00){\usebox{\plotpoint}}
\put(1364.57,471.00){\usebox{\plotpoint}}
\put(1385.33,471.00){\usebox{\plotpoint}}
\put(1406.09,471.00){\usebox{\plotpoint}}
\put(1426.84,471.00){\usebox{\plotpoint}}
\put(1439,471){\usebox{\plotpoint}}
\put(140,557){\usebox{\plotpoint}}
\put(140.00,557.00){\usebox{\plotpoint}}
\put(160.76,557.00){\usebox{\plotpoint}}
\put(181.51,557.00){\usebox{\plotpoint}}
\put(202.27,557.00){\usebox{\plotpoint}}
\put(223.02,557.00){\usebox{\plotpoint}}
\put(243.78,557.00){\usebox{\plotpoint}}
\put(264.53,557.00){\usebox{\plotpoint}}
\put(285.29,557.00){\usebox{\plotpoint}}
\put(306.04,557.00){\usebox{\plotpoint}}
\put(326.80,557.00){\usebox{\plotpoint}}
\put(347.55,557.00){\usebox{\plotpoint}}
\put(368.31,557.00){\usebox{\plotpoint}}
\put(389.07,557.00){\usebox{\plotpoint}}
\put(409.82,557.00){\usebox{\plotpoint}}
\put(430.58,557.00){\usebox{\plotpoint}}
\put(451.33,557.00){\usebox{\plotpoint}}
\put(472.09,557.00){\usebox{\plotpoint}}
\put(492.84,557.00){\usebox{\plotpoint}}
\put(513.60,557.00){\usebox{\plotpoint}}
\put(534.35,557.00){\usebox{\plotpoint}}
\put(555.11,557.00){\usebox{\plotpoint}}
\put(575.87,557.00){\usebox{\plotpoint}}
\put(596.62,557.00){\usebox{\plotpoint}}
\put(617.38,557.00){\usebox{\plotpoint}}
\put(638.13,557.00){\usebox{\plotpoint}}
\put(658.89,557.00){\usebox{\plotpoint}}
\put(679.64,557.00){\usebox{\plotpoint}}
\put(700.40,557.00){\usebox{\plotpoint}}
\put(721.15,557.00){\usebox{\plotpoint}}
\put(741.91,557.00){\usebox{\plotpoint}}
\put(762.66,557.00){\usebox{\plotpoint}}
\put(783.42,557.00){\usebox{\plotpoint}}
\put(804.18,557.00){\usebox{\plotpoint}}
\put(824.93,557.00){\usebox{\plotpoint}}
\put(845.69,557.00){\usebox{\plotpoint}}
\put(866.44,557.00){\usebox{\plotpoint}}
\put(887.20,557.00){\usebox{\plotpoint}}
\put(907.95,557.00){\usebox{\plotpoint}}
\put(928.71,557.00){\usebox{\plotpoint}}
\put(949.46,557.00){\usebox{\plotpoint}}
\put(970.22,557.00){\usebox{\plotpoint}}
\put(990.98,557.00){\usebox{\plotpoint}}
\put(1011.73,557.00){\usebox{\plotpoint}}
\put(1032.49,557.00){\usebox{\plotpoint}}
\put(1053.24,557.00){\usebox{\plotpoint}}
\put(1074.00,557.00){\usebox{\plotpoint}}
\put(1094.75,557.00){\usebox{\plotpoint}}
\put(1115.51,557.00){\usebox{\plotpoint}}
\put(1136.26,557.00){\usebox{\plotpoint}}
\put(1157.02,557.00){\usebox{\plotpoint}}
\put(1177.77,557.00){\usebox{\plotpoint}}
\put(1198.53,557.00){\usebox{\plotpoint}}
\put(1219.29,557.00){\usebox{\plotpoint}}
\put(1240.04,557.00){\usebox{\plotpoint}}
\put(1260.80,557.00){\usebox{\plotpoint}}
\put(1281.55,557.00){\usebox{\plotpoint}}
\put(1302.31,557.00){\usebox{\plotpoint}}
\put(1323.06,557.00){\usebox{\plotpoint}}
\put(1343.82,557.00){\usebox{\plotpoint}}
\put(1364.57,557.00){\usebox{\plotpoint}}
\put(1385.33,557.00){\usebox{\plotpoint}}
\put(1406.09,557.00){\usebox{\plotpoint}}
\put(1426.84,557.00){\usebox{\plotpoint}}
\put(1439,557){\usebox{\plotpoint}}
\put(140,644){\usebox{\plotpoint}}
\put(140.00,644.00){\usebox{\plotpoint}}
\put(160.76,644.00){\usebox{\plotpoint}}
\put(181.51,644.00){\usebox{\plotpoint}}
\put(202.27,644.00){\usebox{\plotpoint}}
\put(223.02,644.00){\usebox{\plotpoint}}
\put(243.78,644.00){\usebox{\plotpoint}}
\put(264.53,644.00){\usebox{\plotpoint}}
\put(285.29,644.00){\usebox{\plotpoint}}
\put(306.04,644.00){\usebox{\plotpoint}}
\put(326.80,644.00){\usebox{\plotpoint}}
\put(347.55,644.00){\usebox{\plotpoint}}
\put(368.31,644.00){\usebox{\plotpoint}}
\put(389.07,644.00){\usebox{\plotpoint}}
\put(409.82,644.00){\usebox{\plotpoint}}
\put(430.58,644.00){\usebox{\plotpoint}}
\put(451.33,644.00){\usebox{\plotpoint}}
\put(472.09,644.00){\usebox{\plotpoint}}
\put(492.84,644.00){\usebox{\plotpoint}}
\put(513.60,644.00){\usebox{\plotpoint}}
\put(534.35,644.00){\usebox{\plotpoint}}
\put(555.11,644.00){\usebox{\plotpoint}}
\put(575.87,644.00){\usebox{\plotpoint}}
\put(596.62,644.00){\usebox{\plotpoint}}
\put(617.38,644.00){\usebox{\plotpoint}}
\put(638.13,644.00){\usebox{\plotpoint}}
\put(658.89,644.00){\usebox{\plotpoint}}
\put(679.64,644.00){\usebox{\plotpoint}}
\put(700.40,644.00){\usebox{\plotpoint}}
\put(721.15,644.00){\usebox{\plotpoint}}
\put(741.91,644.00){\usebox{\plotpoint}}
\put(762.66,644.00){\usebox{\plotpoint}}
\put(783.42,644.00){\usebox{\plotpoint}}
\put(804.18,644.00){\usebox{\plotpoint}}
\put(824.93,644.00){\usebox{\plotpoint}}
\put(845.69,644.00){\usebox{\plotpoint}}
\put(866.44,644.00){\usebox{\plotpoint}}
\put(887.20,644.00){\usebox{\plotpoint}}
\put(907.95,644.00){\usebox{\plotpoint}}
\put(928.71,644.00){\usebox{\plotpoint}}
\put(949.46,644.00){\usebox{\plotpoint}}
\put(970.22,644.00){\usebox{\plotpoint}}
\put(990.98,644.00){\usebox{\plotpoint}}
\put(1011.73,644.00){\usebox{\plotpoint}}
\put(1032.49,644.00){\usebox{\plotpoint}}
\put(1053.24,644.00){\usebox{\plotpoint}}
\put(1074.00,644.00){\usebox{\plotpoint}}
\put(1094.75,644.00){\usebox{\plotpoint}}
\put(1115.51,644.00){\usebox{\plotpoint}}
\put(1136.26,644.00){\usebox{\plotpoint}}
\put(1157.02,644.00){\usebox{\plotpoint}}
\put(1177.77,644.00){\usebox{\plotpoint}}
\put(1198.53,644.00){\usebox{\plotpoint}}
\put(1219.29,644.00){\usebox{\plotpoint}}
\put(1240.04,644.00){\usebox{\plotpoint}}
\put(1260.80,644.00){\usebox{\plotpoint}}
\put(1281.55,644.00){\usebox{\plotpoint}}
\put(1302.31,644.00){\usebox{\plotpoint}}
\put(1323.06,644.00){\usebox{\plotpoint}}
\put(1343.82,644.00){\usebox{\plotpoint}}
\put(1364.57,644.00){\usebox{\plotpoint}}
\put(1385.33,644.00){\usebox{\plotpoint}}
\put(1406.09,644.00){\usebox{\plotpoint}}
\put(1426.84,644.00){\usebox{\plotpoint}}
\put(1439,644){\usebox{\plotpoint}}
\put(140,730){\usebox{\plotpoint}}
\put(140.00,730.00){\usebox{\plotpoint}}
\put(160.76,730.00){\usebox{\plotpoint}}
\put(181.51,730.00){\usebox{\plotpoint}}
\put(202.27,730.00){\usebox{\plotpoint}}
\put(223.02,730.00){\usebox{\plotpoint}}
\put(243.78,730.00){\usebox{\plotpoint}}
\put(264.53,730.00){\usebox{\plotpoint}}
\put(285.29,730.00){\usebox{\plotpoint}}
\put(306.04,730.00){\usebox{\plotpoint}}
\put(326.80,730.00){\usebox{\plotpoint}}
\put(347.55,730.00){\usebox{\plotpoint}}
\put(368.31,730.00){\usebox{\plotpoint}}
\put(389.07,730.00){\usebox{\plotpoint}}
\put(409.82,730.00){\usebox{\plotpoint}}
\put(430.58,730.00){\usebox{\plotpoint}}
\put(451.33,730.00){\usebox{\plotpoint}}
\put(472.09,730.00){\usebox{\plotpoint}}
\put(492.84,730.00){\usebox{\plotpoint}}
\put(513.60,730.00){\usebox{\plotpoint}}
\put(534.35,730.00){\usebox{\plotpoint}}
\put(555.11,730.00){\usebox{\plotpoint}}
\put(575.87,730.00){\usebox{\plotpoint}}
\put(596.62,730.00){\usebox{\plotpoint}}
\put(617.38,730.00){\usebox{\plotpoint}}
\put(638.13,730.00){\usebox{\plotpoint}}
\put(658.89,730.00){\usebox{\plotpoint}}
\put(679.64,730.00){\usebox{\plotpoint}}
\put(700.40,730.00){\usebox{\plotpoint}}
\put(721.15,730.00){\usebox{\plotpoint}}
\put(741.91,730.00){\usebox{\plotpoint}}
\put(762.66,730.00){\usebox{\plotpoint}}
\put(783.42,730.00){\usebox{\plotpoint}}
\put(804.18,730.00){\usebox{\plotpoint}}
\put(824.93,730.00){\usebox{\plotpoint}}
\put(845.69,730.00){\usebox{\plotpoint}}
\put(866.44,730.00){\usebox{\plotpoint}}
\put(887.20,730.00){\usebox{\plotpoint}}
\put(907.95,730.00){\usebox{\plotpoint}}
\put(928.71,730.00){\usebox{\plotpoint}}
\put(949.46,730.00){\usebox{\plotpoint}}
\put(970.22,730.00){\usebox{\plotpoint}}
\put(990.98,730.00){\usebox{\plotpoint}}
\put(1011.73,730.00){\usebox{\plotpoint}}
\put(1032.49,730.00){\usebox{\plotpoint}}
\put(1053.24,730.00){\usebox{\plotpoint}}
\put(1074.00,730.00){\usebox{\plotpoint}}
\put(1094.75,730.00){\usebox{\plotpoint}}
\put(1115.51,730.00){\usebox{\plotpoint}}
\put(1136.26,730.00){\usebox{\plotpoint}}
\put(1157.02,730.00){\usebox{\plotpoint}}
\put(1177.77,730.00){\usebox{\plotpoint}}
\put(1198.53,730.00){\usebox{\plotpoint}}
\put(1219.29,730.00){\usebox{\plotpoint}}
\put(1240.04,730.00){\usebox{\plotpoint}}
\put(1260.80,730.00){\usebox{\plotpoint}}
\put(1281.55,730.00){\usebox{\plotpoint}}
\put(1302.31,730.00){\usebox{\plotpoint}}
\put(1323.06,730.00){\usebox{\plotpoint}}
\put(1343.82,730.00){\usebox{\plotpoint}}
\put(1364.57,730.00){\usebox{\plotpoint}}
\put(1385.33,730.00){\usebox{\plotpoint}}
\put(1406.09,730.00){\usebox{\plotpoint}}
\put(1426.84,730.00){\usebox{\plotpoint}}
\put(1439,730){\usebox{\plotpoint}}
\put(140,817){\usebox{\plotpoint}}
\put(140.00,817.00){\usebox{\plotpoint}}
\put(160.76,817.00){\usebox{\plotpoint}}
\put(181.51,817.00){\usebox{\plotpoint}}
\put(202.27,817.00){\usebox{\plotpoint}}
\put(223.02,817.00){\usebox{\plotpoint}}
\put(243.78,817.00){\usebox{\plotpoint}}
\put(264.53,817.00){\usebox{\plotpoint}}
\put(285.29,817.00){\usebox{\plotpoint}}
\put(306.04,817.00){\usebox{\plotpoint}}
\put(326.80,817.00){\usebox{\plotpoint}}
\put(347.55,817.00){\usebox{\plotpoint}}
\put(368.31,817.00){\usebox{\plotpoint}}
\put(389.07,817.00){\usebox{\plotpoint}}
\put(409.82,817.00){\usebox{\plotpoint}}
\put(430.58,817.00){\usebox{\plotpoint}}
\put(451.33,817.00){\usebox{\plotpoint}}
\put(472.09,817.00){\usebox{\plotpoint}}
\put(492.84,817.00){\usebox{\plotpoint}}
\put(513.60,817.00){\usebox{\plotpoint}}
\put(534.35,817.00){\usebox{\plotpoint}}
\put(555.11,817.00){\usebox{\plotpoint}}
\put(575.87,817.00){\usebox{\plotpoint}}
\put(596.62,817.00){\usebox{\plotpoint}}
\put(617.38,817.00){\usebox{\plotpoint}}
\put(638.13,817.00){\usebox{\plotpoint}}
\put(658.89,817.00){\usebox{\plotpoint}}
\put(679.64,817.00){\usebox{\plotpoint}}
\put(700.40,817.00){\usebox{\plotpoint}}
\put(721.15,817.00){\usebox{\plotpoint}}
\put(741.91,817.00){\usebox{\plotpoint}}
\put(762.66,817.00){\usebox{\plotpoint}}
\put(783.42,817.00){\usebox{\plotpoint}}
\put(804.18,817.00){\usebox{\plotpoint}}
\put(824.93,817.00){\usebox{\plotpoint}}
\put(845.69,817.00){\usebox{\plotpoint}}
\put(866.44,817.00){\usebox{\plotpoint}}
\put(887.20,817.00){\usebox{\plotpoint}}
\put(907.95,817.00){\usebox{\plotpoint}}
\put(928.71,817.00){\usebox{\plotpoint}}
\put(949.46,817.00){\usebox{\plotpoint}}
\put(970.22,817.00){\usebox{\plotpoint}}
\put(990.98,817.00){\usebox{\plotpoint}}
\put(1011.73,817.00){\usebox{\plotpoint}}
\put(1032.49,817.00){\usebox{\plotpoint}}
\put(1053.24,817.00){\usebox{\plotpoint}}
\put(1074.00,817.00){\usebox{\plotpoint}}
\put(1094.75,817.00){\usebox{\plotpoint}}
\put(1115.51,817.00){\usebox{\plotpoint}}
\put(1136.26,817.00){\usebox{\plotpoint}}
\put(1157.02,817.00){\usebox{\plotpoint}}
\put(1177.77,817.00){\usebox{\plotpoint}}
\put(1198.53,817.00){\usebox{\plotpoint}}
\put(1219.29,817.00){\usebox{\plotpoint}}
\put(1240.04,817.00){\usebox{\plotpoint}}
\put(1260.80,817.00){\usebox{\plotpoint}}
\put(1281.55,817.00){\usebox{\plotpoint}}
\put(1302.31,817.00){\usebox{\plotpoint}}
\put(1323.06,817.00){\usebox{\plotpoint}}
\put(1343.82,817.00){\usebox{\plotpoint}}
\put(1364.57,817.00){\usebox{\plotpoint}}
\put(1385.33,817.00){\usebox{\plotpoint}}
\put(1406.09,817.00){\usebox{\plotpoint}}
\put(1426.84,817.00){\usebox{\plotpoint}}
\put(1439,817){\usebox{\plotpoint}}
\put(-8,800){\large{$h_\Delta$}}
\put(1500,0){\large{$\frac{x_2}{\Delta}$}}
\end{picture}

\begin{center}
  {\bf Figure 1: A stripe configuration of charge $9$}
\end{center}

It is interesting to realize that a proper understanding of the
relation between the height of the stripe and the number of zero modes
{\em requires\/} the use of the `smoothed' version of the steplike
scalar field configuration. This should be evident from Figure 1, if
one imagines the curve to be deformed to a steplike configuration.

We remark that the zero modes produced in this way are massless Dirac
fermions in $1+1$ dimensions, because each flavour produces a given
chirality, and the signs are opposite.

\section{Derivative expansion for the fermionic determinant}\label{derexp}
We shall now consider the evaluation of the fermionic determinant, as a
functional of $a_\mu$ and $\varphi$, in a derivative expansion
approximation. The field $a_\mu$ denotes the perturbative part of
the gauge field.

To avoid any risk of confusion, we shall, from now on,
use two-component fermions only. The `effective action'
$\Gamma_F[\varphi, a_\mu]$ is defined as the result of functionally
integrating out the fermionic fields. Using  the Fourier decomposition
for the $x_3=s$ coordinate, we see that
$$
e^{-\Gamma_F[\varphi, a_\mu]} \;=\; \int \prod_{n=-\infty}^{+\infty}
{\mathcal D}{\bar\chi}_n^{(1)} {\mathcal D}\chi_n^{(1)}
{\mathcal D}{\bar\chi}_n^{(2)} {\mathcal D}\chi_n^{(2)}
$$
\begin{equation}\label{defdet1}
\times \,\exp \left\{- \sum_{-\infty}^{+\infty}
[ {\bar\chi}_n^{(1)}(\not \! d + \omega_n + e \varphi)\chi_n^{(1)}
+{\bar\chi}_n^{(2)}(\not \! d - \omega_n - e \varphi)\chi_n^{(2)}
]
\right\} \;.
\end{equation}
Thus $\Gamma_F$ may be written in terms of  fermionic determinants in
$2+1$ dimensions:
$$
\Gamma_F[\varphi, a_\mu]\,=\,-\sum_{-\infty}^{+\infty}
\left\{
\ln \det (\not \! d + \omega_n + e \varphi)
+\ln \det (\not \! d - \omega_n - e \varphi)]
\right\}
$$
\begin{equation}\label{fullgf}
= -\sum_{-\infty}^{+\infty}
{\rm Tr} \ln \left[-\not\! d^2 + (\omega_n + e \varphi)^2\right]\;.
\end{equation}

Large gauge transformations wind up a number of times around the
periodic coordinate, $s$. Then they correspond to constant shifts
in the scalar field $\varphi$:
\begin{equation}\label{large}
\varphi(x) \;\to\; \varphi(x) \,+\, \frac{2\pi n}{e L} \;,
\end{equation}
where $n$ is an integer. It is now clear, from the general
expression (\ref{fullgf}), that {\em all\/} the Fourier modes must
be kept, if invariance under (\ref{large}) is to be maintained.

In the derivative expansion technique, which we shall use in order to
evaluate $\Gamma_F$, the leading, zero derivative term, can usually  be
treated exactly. This term may be thought of as depending basically on
the constant component of the fields. The following terms in the
derivative expansion depend of course also on the fluctuating part,
which is assumed to be small in comparison with the constant part.

The piece of the effective action depending on the constant part is
usually regarded as an `effective potential'. In our case, there is no
point in keeping a constant part for $a_\mu$ since, if present, it
could be gauged away (because the `planar' system is assumed to have a
trivial topology). From (\ref{large}), it is evident that to keep a
constant component for $\varphi$ is, in turn, crucial.  We thus
decompose  $\varphi$ into two  pieces,

\begin{equation}\label{phsplt}
\varphi (x)\;=\; \varphi_0 \,+\, {\tilde \varphi} (x) \;,
\end{equation}
where the constant $\varphi_0\,=\,\langle \varphi(x)\rangle$ denotes
the ($x_\mu$) spacetime average of $\varphi (x)$, and
${\tilde\varphi}(x)\,=\,\varphi(x)\,-\,\varphi_0$. Then we factorize a
constant field determinant,
\begin{equation}
\Gamma_F[\varphi, a_\mu] \;=\; \Gamma_F[\varphi_0, 0]\,+\, \Delta
\Gamma_F[{\tilde\varphi}, a_\mu]
\end{equation}
where
\begin{equation}\label{effpot}
\Gamma_F[\varphi_0,0] \,\equiv\, \int d^3x V_{eff}(\varphi)
\,=\,-\sum_{-\infty}^{+\infty}
{\rm Tr}\ln \left[-\not\! \partial^2 + (\omega_n + e \varphi_0)^2
\right]
\end{equation}
and
\begin{eqnarray}\label{defdelta}
\Delta \Gamma [{\tilde\varphi},a_\mu]\;&=&\;
-\,\sum_{n=-\infty}^{+\infty} {\rm Tr} \ln \left[ 1 \,+\,e
(\not\!\partial \,+\omega_n +\,e\varphi_0)^{-1} (i \not \!a\,+\,
{\tilde \varphi}) \right] \; \nonumber \\
&-&\,\sum_{n=-\infty}^{+\infty} {\rm Tr} \ln \left[ 1 \,+\,e
(\not\!\partial \,-\omega_n -\,e\varphi_0)^{-1} (i \not \!a\,-\,
{\tilde \varphi}) \right] \; .
\end{eqnarray}
We have introduced the notation `$V_{eff}$' under the integral
symbol in (\ref{effpot}), to emphasize the property that it will
play the role  of an effective potential for $\varphi$. At the end
of the evaluation, we shall follow the common practice of
replacing $\varphi_0$ by $\varphi$ in $V_{eff}$. This
approximation is justified within the derivative expansion
technique, as long as the hypothesis leading to that expansion are
valid~\cite{aitf,sem1}.

Large gauge invariance means that this potential is a  {\em
periodic function\/} of $\varphi$. The constant field determinant
is  evaluated by taking the functional and Dirac traces in
momentum space: $$ \Gamma_F[\varphi_0, 0]\,=\,-
\sum_{n=-\infty}^{n=+\infty}{\rm Tr} \ln \left[ - {\not
\!\partial}^2 + (\omega_n + e \varphi_0)^2 \right] $$
\begin{equation}
=\,-2 \int d^3x\, \sum_{n=-\infty}^{n=+\infty}\int \frac{d^3p}{(2\pi)^3}
\ln \left[p^2 + (\omega_n + e \varphi_0)^2\right] \;.
\end{equation}
To evaluate the sum over $n$, we follow techniques of standard application
in finite temperature quantum field theory~\cite{kapu}. We first take
advantage of the fact that the series runs from $n=-\infty$ to
$n=+\infty$, to write
\begin{eqnarray}
\Gamma_F[\varphi_0, 0]&=&- \int d^3x\,
\sum_{n=-\infty}^{n=+\infty}\int \frac{d^3p}{(2\pi)^3}
\left\{ \ln[p^2 + (\omega_n + e \varphi_0)^2]
 \right.\nonumber\\
&& + \left. \ln[p^2 + (\omega_n - e \varphi_0)^2]
\right\}
\;,
\end{eqnarray}
which is easily rearranged as
\begin{eqnarray}
\Gamma_F[\varphi_0, 0]&=&- \int d^3x\,
\sum_{n=-\infty}^{n=+\infty}\int \frac{d^3p}{(2\pi)^3}
\left\{ \ln[\omega_n^2 + (p + i e \varphi_0)^2]
\right.\nonumber\\
&&  + \left.
\ln[\omega_n^2 + (p - i e \varphi_0)^2]
\right\}
\;.
\label{rear}
\end{eqnarray}
Each one of the terms in (\ref{rear}) may be evaluated by translating
known results about the free energy for a system of free {\em
bosons\/}~\cite{kapu}, what yields,
\begin{equation}
\Gamma_F[\varphi_0, 0]\,=\,- \int d^3x\,
\int \frac{d^3p}{(2\pi)^3}
\left\{ \ln[1 - e^{-L (p + i e \varphi_0)} ]
+
\ln[1 - e^{-L(p - i e \varphi_0}]
\right\}
\;,
\end{equation}
where we have ignored the `zero point' contribution, since it is a
$\varphi$-indepen\-dent constant. We shall, however, fix the ambiguity in
the renormalization of this infinite constant by demanding that the
$V_{eff}$ vanishes at its minima.

The integral over $p$ can be performed, and the result can be presented
as a series,
\begin{equation}
\Gamma_F[\varphi_0, 0]\,=\, \int d^3x\, \frac{4}{\pi^2 L^3}
\sum_{n=-\infty}^{\infty} \frac{1}{n^4} [\cos(n e L \varphi_0) -
(-1)^n ] \;,
\end{equation}
or:
\begin{equation}
\int d^3x V_{eff}\,=\, \int d^3x\,\left[ \frac{4}{\pi^2 L^3}
\sum_{n=-\infty}^{\infty} \frac{1}{n^4} \cos(n e L \varphi_0) -
\frac{14 \pi^2}{5 L^3} \right] \;.
\label{suis}
\end{equation}


Let us next consider the remaining part of the effective action, denoted
$\Delta \Gamma_F[{\tilde\varphi},a]$. Besides a dependence on $\varphi_0$,
it will also depend on ${\tilde\varphi}$ (the fluctuating part of $\varphi$)
and $a_\mu$.

To be consistent with the derivative expansion, and also because
both $\varphi$ and $a_\mu$ are proportional to the same coupling
constant $e$, the expansion must treat both fields as a single
entity~\footnote{Indeed, they originate in different components of
the {\em same\/} $3+1$ dimensional gauge field $A_\alpha$.}. To
this end, we have found illuminating to define a new vector field
$\sigma_\mu$, which summarizes the information on the $a_\mu$ and
${\tilde\varphi}$ configurations, through the relations
\begin{eqnarray}
\epsilon_{\mu\nu\lambda}\partial_\nu \sigma_\lambda &=& e a_\mu
\nonumber\\
\partial_\mu \sigma_\mu &=& e {\tilde \varphi}(x) \;.
\label{defsigma}
\end{eqnarray}
Of course, this involves the assumption that the Lorentz gauge
$\partial \cdot a =0$ has been adopted for the gauge field. The
number of independent components for $\sigma_\mu$ is three, since
this field is, in principle, not constrained by any gauge
invariance requirement, and this matches the number of components
for the transverse $a_\mu$ (two) plus the scalar fields (one).
Using trivial properties of the $\gamma$-matrices in $2+1$
dimensions, we observe that ${\Delta \Gamma}_F
({\tilde\varphi},a)$ may be written as
\begin{eqnarray}\label{delf}
\Delta\Gamma_F[{\tilde\varphi},a_\mu]&=& - \sum_{n=-\infty}^{\infty}
\left\{ {\rm Tr} \ln [1 + (\not \! \partial + \omega_n + e\varphi_0)^{-1}
(\not\!\partial \not\!\sigma)] \right. \nonumber\\
 &+&\left. \varphi \leftrightarrow -\varphi  \right\}
\;.
\end{eqnarray}
Because of its explicit Lorentz covariance, this expression could be
regarded as a consistency check for the procedure of introducing
$\sigma_\mu$ as a Lorentz vector.

Taking into account the fact that (\ref{delf}) is explicitly even
under parity transformations, we may write
\begin{equation}
\Delta\Gamma_F[\varphi, a_\mu] \,=\,
\Delta\Gamma_{{\rm even}}[\partial_\mu \varphi - {\tilde F}_\mu]
\end{equation}
where the `even' label refers to the behaviour under parity transformations,
and \mbox{${\tilde F}_\mu = \epsilon_{\mu\nu\lambda} \partial_\nu a_\lambda$}.
The particular dependence of $\Delta\Gamma_F$ on its arguments is due
to the fact that:
\begin{equation}
\sigma_\mu \,=\, e \frac{1}{\partial^2} (\partial_\mu \varphi
- {\tilde F}_\mu) \;.
\end{equation}

We calculate the leading term in a derivative expansion for
$\Delta\Gamma_F$, by noting that, expanding up to second order in
derivatives the $2+1$ dimensional object
\begin{equation}
\gamma [\sigma_\mu]\,=\,
{\rm Tr} \ln [1 + (\not \! \partial + \omega_n + e\varphi_0)^{-1}
(\not\!\partial \not\!\sigma)] \;,
\end{equation}
we obtain
\begin{equation}
\gamma [\sigma_\mu ] \,=\,\frac{e^2}{24 \pi |\omega_n +
e\varphi_0|} \int d^3x \left[ (\partial_\mu \varphi - {\tilde
F}_\mu) (\partial_\mu \varphi - {\tilde F}_\mu)\right] \;.
\end{equation}
 Then we easily see that
\begin{eqnarray} \label{subl}
\Delta\Gamma_F[\varphi, a_\mu] \,&=& \{ \gamma [\sigma_\mu]+
(\varphi \leftrightarrow - \varphi )\} \nonumber\\ &=& \,
\frac{e^2}{12\pi} \,  \int d^3 x \,\sum_{n=-\infty}^{\infty} \frac{1}{|\omega_n
+ e\varphi_0|} [(\partial_\mu \varphi)^2 + ({\tilde
F}_\mu)^2 ]\;.
\end{eqnarray}

Namely, it corresponds to a local Maxwell term for $a_\mu$ and    a
local kinetic term for $\varphi$.

In summary, the effective action induced by  the integration of the
fermio\-nic degrees of freedom is, to second order in a derivative
expansion:
\begin{eqnarray} \label{efac}
{\Gamma_F}[\varphi, a_\mu] &=&  \int d^3x\,\left[ \frac{4}{\pi^2
L^3} \sum_{n=-\infty}^{\infty} \frac{1}{n^4} \cos(n e L \varphi) -
\frac{14 \pi^2}{5 L^3} \right] \nonumber \\ &+& \,
\frac{e^2}{12\pi} \, \int d^3 x \,
 \sum_{n=-\infty}^{\infty} \frac{1}{|\omega_n
+ e\varphi|} [(\partial_\mu \varphi)^2 + ({\tilde F}_\mu)^2] \;,
\end{eqnarray}
where we have  replaced $\varphi_0$ by $\varphi$ in the
contributions from the effective potential (\ref{suis}) and the
subleading term in the derivative expansion of the fermionic
determinant (\ref{subl}). As already mentioned, this procedure is
justified within the derivative expansion approximation.

\section{Massless $QED_4$}\label{qed}
In previous sections, $A_\alpha$ was just a background for the fermion
fields. Here, we shall incorporate a Maxwell action so that $A_\alpha$
becomes dynamical. We then start from $QED_4$ with the fermionic field
constrained to the region $0\leq x_3\leq L$. The Abelian gauge field
is, in principle, defined on an unconstrained region.  However, its
dynamics is determined entirely by the sources, which are, indeed,
confined to $0\leq x_3\leq L$. When the momenta involved in the
processes are small in comparison with the scale $L^{-1}$, an
effective theory can be obtained for the relevant degrees of freedom,
which describe the dynamics on the plane $x_3=0$.  This effective
model contains Dirac fermions in interaction with a dynamical Abelian
gauge field, and with an (also dynamical) scalar field, in $2+1$
dimensions.  The spectrum of this theory also contains stripes:
(Dirac) fermionic zero modes localized on linear defects.

Our starting point shall be to consider a model defined in terms of
$S$, the action for massless $QED$ in four Euclidean dimensions
\begin{equation}\label{defs}
S \;=\; S_F \,+\, S_G \;,
\end{equation}
where $S_F$ and $S_G$ denote the fermionic and gauge field parts of
the action, respectively. They are given by
\begin{equation}\label{defsf1}
S_F\;=\;\int d^4x\,{\bar\psi}(x)\not\!\! D\,\psi(x) \;\;\;,\;\;\;
\not\!\!D\,=\,\not\!\partial+i e \not\!\!A
\end{equation}
and
\begin{equation}\label{defsg}
S_G \;=\; \int d^4x \,\frac{1}{4} F_{\alpha\beta}F_{\alpha\beta} \;\;,
\;\; F_{\alpha\beta}\,=\, \partial_\alpha A_\beta - \partial_\beta
A_\alpha \;.
\end{equation}
We assume that the gauge field components, $A_\alpha$, are defined
over the full four-dimensional Euclidean spacetime, so that the
integration region in (\ref{defsg}) is unbounded.

The fermionic field is, in turn, defined on a region of small width
$L$ in the third spatial dimension, namely: \mbox{$0\leq x_3 \leq L$},
to simulate the physical situation of fermions in a quasi planar
system. There are, of course, many different choices for the boundary
conditions of the fermionic field at $x_3=0$ and $x_3=L$, all of them
compatible with current conservation. In principle, the most natural
one would be to impose the vanishing of the normal component of the
current on the borders, namely, $j_3 =0$ at $x_3\,=\,0,\,L$.  Although
this condition is natural, it is not an easy one to deal with from the
point of view of the calculation, because it explicitly breaks
translation invariance along $x_3$.  To avoid this technical
inconvenience, we prefer to use the simpler assumption that the
fermionic field is $L$-periodic in $x_3$.  This allows us to Fourier
expand in that coordinate.  From the point of view of the effective
physics in the \mbox{$0 \leq x_3 \leq L$} region (the `dimensionally
reduced' theory), both choices should lead to qualitatively similar
results. The important point is, as we shall see, that the existence
of a finite dimension allows for a description in terms of an infinite
number of fermionic modes in the dimensionally reduced spacetime. The
precise nature of the boundary conditions will of course affect the
details of this phenomenon, like the spacing between the fermionic
modes, but not the gross features and properties of the system.

The properties and objects we shall be concerned with can all be
obtained, in principle, from the knowledge of ${\mathcal
  Z}[j;{\bar\eta},\eta]$, the generating functional of Green's
functions
\begin{equation}
{\mathcal Z}[j;{\bar\eta},\eta]\,=\, \int {\mathcal D}A_\mu
{\mathcal D}{\bar\psi} {\mathcal D}\psi \,
\exp \left[- S + \int d^3 x \int_0^L ds (j_\alpha A_\alpha +
{\bar\eta}\psi + {\bar\psi}\eta )\right]
\end{equation}
with $S$ as defined in~(\ref{defs}).  As we are interested exclusively
in the phenomena localized on the $x_3=0$ plane, both the external
source $j_\alpha$, and the fermionic current \mbox{$i e
  {\bar\psi}\gamma_\alpha\psi$} are confined to the \mbox{$0 \leq s
  \leq L$} region.

It is evident that, {\em from the point of view of the description of
  the physics on the $x_3=0$ plane\/}, the action for the gauge field
yields more information than what is actually needed in our situation,
since we are not going to consider processes with sources outside the
region $0 \leq x_3 \leq L$. It is possible, as we shall see, to obtain
a $2+1$ dimensional action describing precisely the planar dynamics,
and containing fermion and gauge fields with support on the $x_3=0$
plane. In a sense, we are going to `integrate out' gauge field modes,
corresponding to excitations lying outside the plane.

As a first step, let us see how to pass to a description where the
relevant gauge field dynamics is determined by an effective $2+1$
dimensional gauge field action.  We may write $S_G^J$, the part of the
action which involves the gauge field and its source, as follows:
\begin{equation}\label{sgj}
S_G^J[A,J] \;=\; S_G[A] \,+\,S_{gf}[A] \,+\, \int d^4x \, J_\alpha(x)
A_\alpha (x)\,
\end{equation}
where we have included a gauge fixing term, $S_{gf}[A]$, which, for
the sake of simplicity, we take to be of the Feynman
type~\footnote{The following derivation can, of course, also be
  implemented for different gauge fixings, but the calculations become
  more involved.}.
\begin{equation}
S_{gf}\,=\, \int d^4x \, \frac{1}{2} (\partial \cdot A)^2 \;.
\end{equation}

$J_\alpha$ stands for the `full' current, namely, the external source
for the gauge field $j_\alpha$ plus the fermionic current \mbox{$i e
  {\bar \psi}(x) \gamma_\alpha \psi (x)$}. We then integrate out the
gauge field, obtaining a non-local current-current interaction:
\begin{equation}\label{nlint}
\int {\mathcal D} A_\alpha \, \exp \{- S_G^J[A,J] \}\;=\; \exp \{ -
S_{nl}[J] \}
\end{equation}
where
\begin{equation}\label{snl}
S_{nl}[J] \;=\; -\frac{1}{2}\int d^4x \,d^4y \,J_\alpha(x)
\,K_{\alpha \beta}(x-y) \,J_\beta (y)
\end{equation}
with $K_{\alpha \beta}(x-y)$ denoting the (Feynman gauge) gauge field
propagator:
\begin{equation}\label{defk}
K_{\alpha \beta}(x-y)\;=\; \int \frac{d^4k}{(2\pi)^4} \,
e^{i k \cdot (x-y)} \frac{\delta_{\alpha\beta}}{k^2}\;.
\end{equation}
Taking now into account the fact that the current $J_\alpha$ vanishes
when either $x_3 < 0$ or $x_3 > L$, we see that (\ref{snl}) may be
written more precisely as follows:
\begin{equation}\label{nlint1}
S_{nl}[J] =-\frac{1}{2}\int d^3x \int_0^L ds_1\int d^3y \int_0^L
ds_2\, J_\alpha(x,s_1)K_{\alpha \beta}(x-y,s_1-s_2) J_\beta (y,s_2) \,.
\end{equation}
In the $L \rightarrow 0$ limit, the current may be taken to be
approximately $x_3$ independent, at least from the effective theory
point of view, since a dependence on $x_3$ in this length scale would
correspond to a momentum component comparable to the scale $L^{-1}$.
Thus, we can replace in (\ref{nlint1}) $J_\alpha(x,s)$ by its average
${\bar J}_\alpha(x)$, defined by
\begin{equation}\label{defjb}
{\bar J}_\alpha(x)\;=\; \frac{1}{L}\, \int_0^L ds J_\alpha(x,s)
\;.
\end{equation}

On the other hand, regarding the propagator, as $0\leq s_{1,2}\leq L$,
we may use the approximation
\begin{equation}
K_{\alpha \beta}(x-y,s_1-s_2) \;\simeq\;{\bar K}_{\alpha \beta}(x-y)
\end{equation}
where
\begin{equation}
{\bar K}_{\alpha \beta}(x-y)\;=\; K_{\alpha \beta}(x-y,0)\;.
\end{equation}

Therefore, (\ref{snl}) reduces to a non local current-current
interaction ${\bar S}_{nl}$ in $2+1$ dimensions
\begin{equation}
{\bar S}_{nl}[{\bar J}] \;=\; - \frac{L^2}{2} \int d^3x  \int d^3y \,
{\bar J}_\alpha(x) {\bar K}_{\alpha \beta}(x-y)  {\bar J}_\beta (y) \;.
\end{equation}

To obtain a more explicit form for \mbox{${\bar K}_{\alpha
    \beta}(x-y)$}, we may write it in terms of its Fourier
representation
\begin{equation} \label{defk2}
{\bar K}_{\alpha \beta}(x-y) \,=\,
\int {\frac{d^3k}{(2\pi)^3}} \,e^{i k \cdot (x-y)}
\int_{-\infty}^{+\infty} \frac{dk_3}{2\pi} \,
\frac{\delta_{\alpha\beta}}{k^2}\;.
\end{equation}
Integrating over $k_3$, we easily see that the effective
three-dimensional kernel for the currents ${\bar J}_\mu(x)$ is
\begin{equation}\label{eq:kernel3}
{\bar K}_{\mu \nu}(x-y) = \int \frac{d^3k}{(2\pi)^3} \, e^{i k
\cdot (x-y)} \frac{ \delta_{\mu\nu}}{\sqrt{k^2}}\;.
\end{equation}
while for the $x_3$ component we have the scalar ${\bar K}_s$:
\begin{equation}
{\bar K}_s(x-y) \;=\; \int \frac{d^3k}{(2\pi)^3} \,
e^{i k \cdot (x-y)} \frac{1}{\sqrt{k^2}}\;.
\end{equation}
Notice that from now on $k$ denotes the three dimensional momentum
vector. ${\bar K}_s$ is treated separately, because $A_3$ behaves in
fact as a {\em scalar\/} field under spacetime coordinate
transformations in the $x_3=0$ hyperplane.

It is important to realize that the $2+1$ dimensional current ${\bar
  J}_\mu$ is conserved:
\begin{equation}
\partial_\mu {\bar J}_\mu \,=\,0 \;.
\end{equation}
This follows from the fact that the current is approximately
$x_3$-independent, in particular: $\partial_3 J_3=0$, plus the usual
continuity equation $\partial_\alpha J_\alpha=0$.  Then we may
actually replace $\delta_{\mu\nu}$ in (\ref{eq:kernel3}) by its
transverse part:
\begin{equation}\label{deltr}
\delta^{\perp}_{\mu\nu}\;=\; \delta_{\mu\nu}-\frac{k_\mu k_\nu}{k^2}\;.
\end{equation}
Summarizing, the result of integrating out the gauge field may, in the
small-$L$ limit, be represented in terms of an effective $2+1$
dimensional action
$$
{\bar S}_{nl}[{\bar J}] \;=\; \frac{L^2}{2} \int d^3x \int d^3y \,
{\bar J}_\mu(x) {\bar K}^{\perp}_{\mu \nu}(x-y){\bar J}_\nu (y)
$$
\begin{equation}
+ \frac{L^2}{2} \int d^3x  \int d^3y {\bar J}_s(x) {\bar K}_s (x-y)
{\bar J}_s (y)
\end{equation}
with ${\bar K}^{\perp}_{\mu\nu}$ as in (\ref{eq:kernel3}), but with
$\delta^{\perp}_{\mu\nu}$ instead of $\delta_{\mu\nu}$.

This effective action ${\bar S}_{nl}[{\bar J}]$ for the currents, can
be equivalently rewritten as arising from the integration of a gauge
field $a_\mu$, and a scalar field $\varphi$, both defined in the three
dimensional space-time $x_3 =0$. In other words, we can express the
above action as the result of functionally integrating out (in $2+1$
dimensions) a gauge field $a_\mu$ and a scalar field $\varphi$, with
the following action:
$$
S'_{nl}[{\bar J};a_\mu,\varphi] \;=\; \frac{1}{2}\int d^3x \int
d^3y \; a_\mu(x) \; {\bar K}^{\perp -1}_{\mu \nu}(x-y)\; a_\nu (y)
$$
$$
+\,\frac{1}{2}\;\int d^3x\int d^3y \;\varphi(x)\; {\bar
  K}_s^{-1}(x-y) \; \varphi(y)
$$
\begin{equation}
+\, \int d^3x  \; {\bar J}'_\mu(x) a_\mu(x) \,+\,
\int d^3 x \; {\bar J}'_s(x)\varphi(x) \;,
\end{equation}
where ${\bar J}'_\alpha (x) = L {\bar J}_\alpha (x)$. Note that the
inclusion of this factor $L$ in the definition of the current reduces
by one its mass dimensions, as it corresponds to the transition from
fermionic fields in $3+1$ dimensions to $2+1$ dimensions. The bosonic
fields, $a_\mu$ and $\varphi$, have the same mass dimension than their
$3+1$ dimensional counterparts, because they are equipped with
unusual, non-local kinetic terms.

Because of current conservation, we may also write a gauge invariant
form for $S'_{nl}[{\bar J};a_\mu,\varphi]$, by introducing the
longitudinal part of $a_\mu$ into the game, obtaining:
$$
S'_{nl}[{\bar J};a_\mu,\varphi] \;=\; \frac{1}{4}\int d^3x
F_{\mu\nu}\frac{1}{\sqrt{-\partial^2}}F_{\mu\nu} \;+\; \int d^3x \;
{\bar J}'_\mu(x) a_\mu(x)
$$
\begin{equation}
  \;+\; \int d^3 x \;
{\bar J}'_s(x)\varphi(x) \; \label{sgau} \;+\;\frac{1}{2}\int
d^3x \partial_\mu\varphi \frac{1}{\sqrt{-\partial^2}} \partial_\mu
\varphi
\label{dosl}
\end{equation}
where $F_{\mu\nu}= \partial_\mu a_\nu - \partial_\nu a_\mu$. This
gauge invariant form should, of course, be gauge fixed in order to
recover a regular action.

We have thus re-derived a known result: the electromagnetic
interaction due to a Maxwell action, if restricted to charges
living on a planar section of space, can be reproduced by
introducing a non-local gauge invariant action in $2+1$
dimensions, like in~\cite{M1,M2}.  This is of course different to
the usual, Kaluza-Klein like `dimensional reduction' prescription
and in fact is closer to an alternative prescription proposed in
\cite{G}. To understand the basic differences between the two
prescriptions, let us concentrate in the case of a scalar field in
$3+1$ dimensions with action
\begin{equation}
S^{(3+1)
} = \frac{1}{2}\int d^4x \partial_\mu \varphi\partial^\mu\varphi
\label{ss}
\end{equation}
with propagator $K^{3+1}$ satisfying in coordinate space the usual
Green function equation
\begin{equation}
-\partial^2{K}^{(3+1)}(x) = \delta(x)
\label{box}
\end{equation}
so that in coordinate and momentum (Minkowski) space one has,
\begin{eqnarray}
K^{(3+1)}(x) &=& -\frac{4\pi^2i}{\left(x^0\right)^2 -  \left(x^1\right)^2
-\left(x^2\right)^2 - \left(x^3\right)^2 - i0}
\label{a}\\
  \tilde K^{(3+1)}(k) &=&   -\frac{1}{\left(k_0\right)^2 -
 \left(k_1\right)^2
-\left(k_2\right)^2 - \left(k_3\right)^2 + i0}
\label{b}
\end{eqnarray}
The usual Kaluza-Klein dimensional reduction consists in this case in
dropping $k_3^2$ in eq.(\ref{b}). Then, the propagator
$K_{KK}^{(2+1)}$ in the reduced space obeys the analogous to
(\ref{box}),
\begin{eqnarray}
-\partial^2{K}_{KK}^{(2+1)}(x) = \delta(x)
\label{rK}
\end{eqnarray}
so that the reduced scalar theory corresponds to an action
\begin{equation}
S^{(2+1)}_{KK} = \frac{1}{2}\int d^3x \partial_\mu \varphi\partial^\mu\varphi
\label{ssr}
\end{equation}
The alternative prescription which leads to the reduction
presented here corresponds to dropping $x^3$ in (\ref{a}). To see
this, note that if one puts $x_3=0$, the propagator $K^{(2+1)}$
does not satisfy anymore (\ref{rK}) but instead, as can be easily
seen,
\begin{eqnarray}
\sqrt{-\partial^2} {K}^{(2+1)}(x) = \delta(x)
\label{rKn}
\end{eqnarray}
so that the dynamics of the reduced scalar theory is governed by the
action
 \begin{equation}
S^{(2+1)}  = \frac{1}{2}\int d^3x \partial_\mu \varphi
\frac{1}{\sqrt{-\partial^2}}
\partial^\mu\varphi
\label{ssrr}
\end{equation}
which is precisely the non-local action for the scalar field in
(\ref{sgau}).  An analogous result, coinciding again with that in
(\ref{sgau}) can be found for a vector field like $a_\mu$.

Let us note that in the case in which the dimensional reductions
is justified by the fact that the physical system is (quasi)
planar, the alternative prescription which consists in dropping
$x^3$, effectively compels the system to evolve in a plane keeping
the nature of the interaction unchanged: for example, if in $3+1$
dimensions electrons interact via a $1/r$ potential, they still
interact through this potential in $2+1$ but with the third
coordinate constrained to be zero.  In contrast, the usual
Kaluza-Klein prescription not only reduces the space but also
changes drastically the nature of the interaction: in the example
above, the electron interaction would change from $1/r$ to $\log
r$.  If planar electrons are supposed to be subjected to $3+1$
interactions even when they are compelled to move in the plane,
the alternative prescriptions then appears to be more physical
than the usual Kaluza-Klein one.

The prize one pays when using the alternative prescription is the
non-locality of the reduced action.  It is interesting to note
that, if looked at from the point of view of Minkowski spacetime,
this non-locality is due to a branch cut singularity starting at
zero momentum. Amusingly enough, this kind of singularity is
entirely analogous to the one that appears when one considers the
vacuum polarization function for massless fermions in $2+1$
dimensions. In that case, the branch cut singularity is of course
due to the possibility of pair-creating fermions {\em on-shell\/}
for any gauge field configuration with a non-zero momentum. The
situation is, however, different to the `dimensional reduction' of
our example, since the singularity here corresponds to modes that
have been integrated out, and are in principle out of the spectrum
of physical states. Thus the model is to be regarded as an open
system, since the unitarity relations corresponding to a closed
system would in turn require the introduction of asymptotic
massless fermionic excitations into the game, like in the case of
(unitary) massless $QED$ in $2+1$ dimensions.

The reduction of the fermionic part of the action proceeds as in
section \ref{free}, while the evaluation of the fermionic
determinant has already been presented in section \ref{derexp}.

\section{Discussion and conclusions }\label{cyd}

 So far we have obtained the
effective action for the gauge fields (\ref{sgau}), after reducing
the original gauge action from $3+1$ to $2+1$ dimensions, and the
one for the fermionic degrees of freedom (eq.( \ref{efac})), after
integrating out the fermions. The final result is:
\begin{eqnarray}
{\cal S}_{eff}[\varphi, a_\mu] &=&  \int d^3x\,\left[
\frac{4}{\pi^2 L^3} \sum_{n=-\infty}^{\infty} \frac{1}{n^4} \cos(n
e L \varphi) - \frac{14 \pi^2}{5 L^3} \right] \nonumber \\ &+& \,
\frac{e^2}{12\pi} \,  \int d^3 x \,
\sum_{n=-\infty}^{\infty} \frac{1}{|\omega_n
+ e\varphi|} [(\partial_\mu \varphi)^2 + \left(F_{\mu\nu}\right)^2
 ] \nonumber \\ &+& \frac{1}{4}\int d^3x
F_{\mu\nu}\frac{1}{\sqrt{-\partial^2}}F_{\mu\nu}
\;+\;\frac{1}{2}\int d^3x \partial_\mu\varphi \frac{1}{\sqrt{-\partial^2}}
\partial_\mu\varphi \;.
\label{acfina}
\end{eqnarray}
We see that the dynamics for the field that provides an `induced' mass
for the fermions, $\varphi$, is given by a non-local interaction term,
and a series of Sine-Gordon like term. It is also through a series
that is a functional of $\varphi$ that this last field couples to
$a_\mu$.

As mentioned above, non-local effective theories for the
electrodynamics of particles moving on a plane were discussed in
\cite{M1}-\cite{bfo}.  In particular, a Lagrangian containing a
non-local Maxwell term plus a coupling to matter current confined
to a plane and coinciding with the first line in Lagrangian
(\ref{dosl}) was obtained in \cite{M2} where it was proven that
this effective $2+1$ non-local Lagrangian was equivalent to one in
which the current term is replaced by a Chern-Simons term. This
last Lagrangian describes precisely the bosonized version of a
$2+1$ massless fermion theory \cite{M1} and, as shown in
\cite{bfo}, this is not a coincidence. Indeed, the functional
approach to bosonization \cite{FS}-\cite{S}  for $2+1$ free Dirac
fermions with mass $m$, leads to a bosonization recipe which is
{\it exact} for the matter current:
 \begin{equation}
 \bar J^\mu(x) \to \frac{1}{\sqrt{4\pi}} \varepsilon^{\mu\nu\alpha}
 \partial_\nu a_\alpha
 \label{bv}
 \end{equation}
 Concerning the bosonic action, it cannot be written explicitly,
 except for certain particular limits. In particular, in the
 vanishing mass limit one can compute exactly the bosonic action
 \cite{bfo} obtaining
 a result which coincides with the first line in eq.(\ref{dosl}) (if
 one uses (\ref{bv}) to bosonize the matter current).  It is worth
 emphasizing here that we have not started from a $2+1$ model but
 rather from massless $QED_4$, and that the reduction mechanism
 explained above brings into play, besides the usual $a_\mu$ field,
 also an scalar field $\varphi$ which has a non-trivial, sine-Gordon
 like dynamics and it is moreover coupled to $a_\mu$. The necessity
 for the existence of this field may be traced back of course to the
 fact that the compactified dimension allows for a certain non-trivial
 structure in that direction, reflected in the large gauge
 transformations of which the scalar field is the subject.

 To finish this discussion we look at  the propagation of fermions
 in a $(a_\mu, \varphi)$ background. In a first approximation, the
 dynamics of these background fields could be regarded as entirely
 determined by the effective action (\ref{acfina}) only, and as a next
 step one should introduce the reaction of the fermionic fields on the
 bosons.  Looking back at the action (\ref{defdet1}) for the fermions
 in $2+1$, we see that for each species $\chi_n^{(1)}$ and
 $\chi_n^{(2)}$, there will be a fermionic zero mode every time $
 \omega_n + e \varphi$ passes through zero, i.e, when $e \varphi=\pm
 {\frac {2n\pi}{L}}$.  These zero modes are produced because of the
 Callan and Harvey mechanism, since there is a change of sign for the
 mass term of a $2+1$ dimensional fermion.  Since we have two
 $2$-component fermions, and their mass terms have opposite signs,
 there shall be two chiral fermions, of opposite chiralities,
 localized around each defect. This is of course equivalent to saying
 that we shall have a massless {\em Dirac\/} fermionic zero mode
 around each defect. The existence of these gapless zero modes has of
 course relevance for the calculation of transport properties.

 If we restrict ourselves to configurations having an approximate
 invariance under translations along one of the spatial coordinates,
 the effective action for the $\varphi$ field will of course be the
 one of a two dimensional field theory with a potential periodic in
 the field, and with an unusual kinetic term. Despite this last
 property, the periodic nature of the potential implies that there is
 room for the existence of a non-trivial topological charge $Q$,
 defined as for the case of the Sine-Gordon model. If the system is
 invariant under translations in $x_2$, then
\begin{equation}
Q\,=\, \frac{e L}{2\pi} \int_{-\infty}^{+\infty}  d x_1 \partial_1
\varphi (x)\;=\; \frac{e L}{2\pi} [\varphi (+\infty) - \varphi (-\infty)] \;.
\end{equation}
A configuration of definite charge $Q\,=\,q$, corresponds to the field
$\varphi$ starting in one of the minima at $x_1=-\infty$ and covering
$q$ periods of the potential before finishing at another minimum of
the potential, at $x_1=+\infty$. We may assume, for the sake of
clarity, that the effective potential is identical to the first
component of the series. Then, there will be exactly $q$ fermionic
zero modes, since they appear each time the potential passes through a
{\em maximum\/}.  This discussion shows that there is an interesting
relation between the topological solitonic charge and the number of
fermionic zero modes.

\vskip 2cm
\underline{Acknowledgements}: C.~D.~F. and A.~L. are supported by
CONICET and  Instituto Balseiro, Argentina.  This work is
supported in part by grants from ANPCYT (PICT 97/0053, 97/2285 and 
98/03-03924), CONICET, and by Fundaci\'on Antorchas, Argentina.


\newpage
\begin{thebibliography}{99}
\bibitem{raja}An excellent reference on the subject is:\\
  R.\ Rajaraman, {\em Solitons and Instantons\/}, Amsterdam, North
  Holland, (1982).
\bibitem{fl} C.\ D.\ Fosco and A.\ L\'opez, Nucl.\
  Phys.\ {\bf B538}, 685 (1999).
\bibitem{callan} C.\ Callan and J.\
  A.\ Harvey, Nucl.\ Phys.\ B {\bf B250}, 427 (1985).
\bibitem{aitf}I.\ J.\ R.\ Aitchison and C.\ M.\ Fraser; Phys.Rev. {\bf
    D31} 2605 (1985).
\bibitem{sem1}G. Grignani, G. Semenoff and P.
  Sodano; Phys.  Rev. {\bf D53} 7157 (1996).
\bibitem{kapu}J. Kapusta, {\it Finite-Temperature Field Theory\/}, Cambridge Univ.
  Press (1993); Appendix A.3 and references therein.
\bibitem{sem2}G. Grignani, G. Semenoff, P. Sodano and O.  Tirkkonen;
Nucl. Phys. {\bf B473} 143 (1996).
\bibitem{frs} C.\ D.\ Fosco, G.\ L.\ Rossini
  and F.\ A.\ Schaposnik; Phys. Rev. Lett. {\bf 79}, 1980 (1997);
  Phys. Rev. {\bf D56}, 6547 (1997) ; Phys. Rev. {\bf D59}, 085012
  (1999).
\bibitem{ffl} C.\ D.\ Fosco, E.\ Fradkin and A.\ L\'opez ,
  Phys.\ Lett.\ {\bf B451}, 31 (1999).
\bibitem{KKS} A.\ R.\ Kavalov,
  I.\ K.\ Kostov, and A.\ K.\ Sedrakyan, Phys.\ Lett.\ {\bf B175}, 331
  (1986).
\bibitem{M1} E.C.~Marino, Phys. Lett. {\bf B263}, 634
  (1991).
\bibitem{M2} E.C.~Marino, Nucl.  {\bf B408}, 551 (1993).
\bibitem{bfo} D.G.~Barci, C.D.~Fosco and L.E.~Oxman, Phys. Lett. {\bf B375}, 267 (1996).
\bibitem{G} J.D.~Edelstein, C.~N\'u\~nez, F.A.~Schaposnik and
  J.J.~Giambiagi, Mod. Phys. Lett. {\bf A11} (1995) 1037.
\bibitem{FS} E.~Fradkin and F.A.~Schaposnik, Phys. Lett. {\bf B338},
  254 (1994).
\bibitem{S} F.A.~Schaposnik, Phys. Lett. {\bf B356}, 39 (1995).
\end{thebibliography}
\end{document}